%% file: manuscript_jpcl.tex
\documentclass[aps,prb,twocolumn,superscriptaddress,showpacs,english]{revtex4-1}

\usepackage[T1]{fontenc}
\usepackage[latin9]{inputenc}
\usepackage{babel}
\usepackage{amsmath}
\usepackage{amssymb}
\usepackage{wasysym}
\usepackage{graphicx}
\usepackage{xcolor}
\usepackage{graphicx}

\usepackage[linktocpage=true,
  colorlinks=true, 
  pdfborder={0 0 0},
  linkcolor=blue,
  citecolor=red,
  filecolor=yellow,
  urlcolor=blue,
  bookmarks,
  pdfauthor={},
]{hyperref}

\usepackage[nohyperlinks,nolist]{acronym}
\newcommand{\mrchem}{{\sc MRChem}}
\newcommand{\fhi}{{\sc FHI-aims}}
\newcommand{\elk}{{\sc ELK}}
\newcommand{\nwchem}{{\sc NWChem}}
\newcommand{\libxc}{{\sc libxc}}
\newcommand{\xcfun}{{\sc xcfun}}

\usepackage[activate={true,nocompatibility},
           final,tracking=true,
           kerning=true,
           spacing=true,
           factor=1100,
           stretch=10,
           shrink=10]{microtype}

\input{newcommands.tex}

\newcommand{\Basel}{Department of Physics, Universit\"at Basel, Klingelbergstr. 82, 4056 Basel, Switzerland}
\newcommand{\Norway}{Centre for Theoretical and Computational Chemistry, Department of Chemistry,\\ UiT - The Arctic University of Norway, N-9037 Troms\o, Norway}
\newcommand{\Duke}{Department of Materials Science and Mechanical Engineering, Duke University,\\ Durham, NC 27708, USA}

\begin{document}

\title{The Elephant in the Room of Density Functional Theory Calculations}

\author{Stig Rune Jensen} \affiliation{\Norway}
\author{Santanu Saha}     \affiliation{\Basel}
\author{Jos\'e A. Flores-Livas} \affiliation{\Basel}
\author{William Huhn} \affiliation{\Duke}
\author{Volker Blum} \affiliation{\Duke}
\author{Stefan Goedecker}       \affiliation{\Basel}
\author{Luca Frediani} \affiliation{\Norway}

\date{\today}

\begin{abstract}
Using multiwavelets, we have obtained total energies and corresponding
atomization energies for the GGA-PBE and hybrid-PBE0 density functionals
for a test set of 211 molecules with an unprecedented and guaranteed
$\mu$Hartree accuracy. These quasi-exact references allow us to quantify
the accuracy of standard all-electron basis sets that are believed to be
highly accurate for molecules, such as Gaussian-type orbitals (GTOs),
all-electron numeric atom-centered orbitals (NAOs) and full-potential
augmented plane wave (APW) methods.
We show that NAOs are able to achieve the so-called chemical accuracy
(1~kcal/mol) for the typical basis set sizes used in applications, for
both total and atomization energies. For GTOs, a triple-zeta quality
basis has mean errors of $\sim$10~kcal/mol in total energies, while
chemical accuracy is almost reached for a quintuple-zeta basis. Due to
systematic error cancellations, atomization energy errors are reduced
by almost an order of magnitude, placing chemical accuracy within reach
also for medium to large GTO bases, albeit with significant outliers.
In order to check the accuracy of the computed densities, we have also
investigated the dipole moments, where in general, only the largest
NAO and GTO bases are able to yield errors below 0.01~Debye.
The observed errors are similar across the different functionals considered here.
\end{abstract}

\maketitle

Electronic structure calculations are nowadays employed by a large and
steadily growing community, spanning condensed matter physics,
physical chemistry, material science, biochemistry and molecular
biology, geophysics and astrophysics. Such a popularity is in large
part due to the development of \ac{DFT} methods,\cite{hohenberg-kohn}
in their \ac{KS} formulation.\cite{kohn-sham}

Although the exact energy functional of \ac{DFT} is unknown, many
approximate functionals offer an excellent compromise between
accuracy and numerical cost, rivaling often the accuracy that can be obtained with 
correlated methods, such as \ac{CCSD}.~\cite{Goerigk:2010kc,Karton:2008ho,Zhao:2005kj}
During the last decades, extensive efforts have been undertaken to provide ever
more accurate approximations to the exact \ac{XC} functional.~\cite{burke} This
quest for higher accuracy is conceptually captured
by John Perdew's Jacob's ladder analogy,~\cite{Perdew:2001}
leading to the heaven of chemical accuracy: errors of 1\,kcal/mol or
less in atomization energies and other energy differences that are of
primary interest in chemistry and solid state physics. Rungs on this
ladder are the \ac{LDA}, the \ac{GGA},~\cite{PhysRevLett.114.053001}
meta-\acp{GGA},~\cite{sun2015strongly} hybrid and double hybrid
functionals.~\cite{QUA:QUA24849} The best modern \ac{XC} functionals 
come fairly close to this target, with errors of a few kcal/mol
on a wide range of energetic properties relative to experiment,
including atomic and molecular
energies, bond energies, excitation and isomerization energies and
reaction barriers, for main-group elements as well as transition
metals and solids.~\cite{peverati2014quest,Head-Gordon_2016}

The closer we get to chemical accuracy, the more important becomes the
identification of errors due to various other, algorithmic
approximations -- basis sets, integration grids and
pseudopotentials,~\cite{singh2006planewaves,transferability,NLCC}
to cite a few -- which can lead to comparable or even larger
errors, but their influence is hard to quantify. The importance of
this issue has recently been highlighted within the solid state community,
with a substantial effort to assess the influence of such approximations on
the accuracy. Lejaeghere~\etal\cite{lejaeghere2016reproducibility}
compared the GGA-PBE\cite{burke} calculated equations of state for 71
elemental crystals from 15 different widely used \ac{DFT} codes,
employing both all-electron methods as well as 40 different
pseudopotential sets. For the equation of state, most \ac{DFT} codes
agree within error bars that are comparable to those of experiment,
irrespective of the basis-set choice: all-electron \acp{NAO},
\ac{APW} methods or plane waves with pseudopotentials.

\ac{APW} methods~\cite{singh2006planewaves} are widely believed to be
highly accurate, but contain several parameters which are difficult to
adjust and which can influence the results in a more or less erratic
way. Hence, the magnitude of the error cannot be rigorously quantified
without an external reference. Similar limitations exist for atomization
energies of molecules obtained with \ac{GTO} and \ac{NAO} basis sets:
both bases cannot be systematically enlarged to achieve completeness in
the $L^2$ sense, and within standardized basis sets, the convergence to
the exact result cannot be achieved. Additionally, for larger systems,
linear dependency issues can limit the ability of these basis sets to
achieve complete convergence.\cite{Moncrieff:2005tf}

The basic mathematical formalism for \ac{KS} \ac{DFT} calculations leads
to a self-consistent three-dimensional partial differential equation. 
What makes the solution of this equation so challenging are the accuracy
requirements for the physically and chemically relevant energy differences. 
For instance, the atomization energy of the largest molecule (SiCl$_4$) in
our data set is less than 1 Ha, but it is computed as a difference of energies
in the order of $\sim$2000 Ha. Hence we need at least 7 correct decimal places
in the total energy of the molecule to get the atomization energy within chemical
accuracy of 1~kcal/mol $\sim$ milli-Hartree (mHa), and even 10 decimal places
for micro-Hartree ($\mu$Ha) accuracy.

For isolated atoms, and using the appropriate numerical techniques,
the associated many-particle problem can be solved with essentially
arbitrary numerical precision. Virtually converged LDA energies for
spherical atoms are available in the NIST data
base.~\cite{kotochigova1997atomic} For a few dimers, highly
accurate energies have been calculated,~\cite{Ballone:1990} and an
attempt to obtain total energies free of basis set error was made also
for general molecular systems,~\cite{Becke:1992} but the accuracy of
this approach seems to be limited to around 1~kcal/mol.
Similar accuracies were achieved for solids 
with semicardinal wavelets.~\cite{arias}

In spite of all the progress that has been made in numerical techniques
to harness the power of quantum mechanical theory and simulations,
none of the traditional techniques is able to furnish, unambiguously,
atomization energies for molecules with arbitrary numerical precision. 
A straightforward, uniform grid- or Fourier transform-based approach
is ruled out since it is impossible to provide sufficient resolution
for the rapidly varying wave functions near the nucleus. Other basis
set techniques are critically hampered by non-orthogonality, which
leads to inevitable algebraic ill-conditioning problems at small but
finite residual precision.~\cite{Moncrieff:2005tf} Because of these
problems a large part of the community resorts to pseudopotentials
methods,~\cite{singh2006planewaves} where the $Z/|r-R|$ potential
is replaced by a smoother potential that retains approximately the
same physical properties of the all-electron atom. The smooth
pseudopotential then allows to obtain arbitrarily high accuracy with
systematic basis sets such as plane waves. The limitation is that the
pseudopotential introduces an approximation error, the magnitude of which
is hard to quantify.

The current \textit{de facto} standard technique to assess errors of
different methods does not rely on an absolute reference: errors are
instead estimated by comparing results obtained with increasingly large
bases.~\cite{NIST,papas2006concerning} The development of Multiwavelet
methods~\cite{Harrison:2003cn,Harrison:2004vua,Yanai:2004tr,Yanai:2004vo}
have fundamentally changed the situation. \acp{MW} are systematic, adaptive,
and can be employed in all-electron calculations. With this approach, it is
now possible to achieve all-electron energies with arbitrarily small errors.

In the present work, we use \acp{MW} to obtain error bars of less than a
$\mu$\ac{Hartree} in the atomization energies for a large test set of 211
molecules with standard \ac{DFT} functionals. We focus on three widely used
and well established functionals, LDA-SVWN5,~\cite{svwn5:1980}
GGA-PBE,~\cite{burke} as well as hybrid-PBE0.~\cite{pbe0adam} PBE and PBE0
are both relatively accurate for atomization energies,~\cite{paier2005perdew}
and have stood the test of time.~\cite{Medvedev:2017dj} Our \ac{MW} results
provide quasi-exact reference values that can be employed to quantify the
accuracy of standard basis sets, such as \ac{GTO}, \ac{NAO}, and \ac{APW}
methods, as well as of novel approaches based for instance on finite element
methods~\cite{pask,Losilla:2012bu,Toivanen:2015hp,doi:10.1021/acs.jctc.6b01207}
or discontinuous Galerkin methods.~\cite{galerkin}
   
%\section{Theory}

Real-space methods have a long history in computational chemistry and
have been used for benchmarking purposes for
decades.~\cite{frediani2015real} However, because of the so-called
\emph{curse of dimensionality}, the na\"ive numerical treatment of
molecular systems is prohibitively expensive, and its applicability
relies on high symmetry to reduce the
dimensionality.~\cite{Sundholm:1994hj} The multi-scale nature of the
problem renders the traditional uniform grid discretization highly
inefficient, unless the problematic nuclear region is treated
separately, e.g. by means of pseudopotentials. The mathematical
theory to solve these issues was developed in the '90s, when Alpert
introduced the \ac{MW} basis, allowing for non-uniform grids
with strict control of the discretization error, as well as sparse
representations of a range of physically important operators, with
high and controllable precision.~\cite{Alpert:1993wd,Alpert:1999hv}

Alpert's construction starts from a standard polynomial basis of
order $k$, such as the Legendre or the Interpolating polynomials,
re-scaled and orthonormalized on the unit interval $[0,1]$. Then,
an orthonormal scaling basis at refinement level $2^{-n}$ is
constructed by dilation and translation of the original basis
functions $\scaling{n}{i,l}(x) = 2^{n/2}\scaling{}{i}(2^n x - l)$,
where \scaling{n}{i,l} is the $i$-th polynomial in the interval
$[l/2^n,(l+1)/2^n)]$ at scale $n$. The set of scaling functions
on all $2^n$ translations at scale $n$ defines the scaling space
$V_k^n$, and in this way a ladder of spaces is constructed such that
the complete $L^2$ limit can be approached in a systematic manner:
\begin{equation}
	V_{k}^{0}\subset V_{k}^{1}\subset\cdots\subset
	V_{k}^{n}\subset\cdots\subset L^2. \label{nest_prop}
\end{equation}
The wavelet spaces $W^{n}_{k}$ are simply the orthogonal complement of
two subsequent scaling spaces $V_{k}^{n}$ and $V_{k}^{n+1}$:
\begin{equation}\label{eq:sum_wav_scal}
	W_{k}^{n} \oplus V_{k}^{n}=V_{k}^{n+1},\qquad W_{k}^{n}\perp V_{k}^{n}.
\end{equation}
Completeness in the $L^2$ sense can be achieved both by increasing the
polynomial order (larger $k$) of the basis and by increasing the
refinement in the ladder of spaces (larger $n$).

Two additional properties are essential in achieving fast and accurate
algorithms: the vanishing moments of the wavelet functions and the disjoint
support of the scaling and wavelet functions. The former leads to fast
convergence in the representation of smooth functions and narrow-banded
operators, whereas the latter enables simple algorithms for adaptive
refinement of the underlying numerical grid, which is essential to limit
storage requirements. The extension to several dimensions is achieved
by standard tensor-product methods; to minimize the impact of the
curse of dimensionality, it is necessary to apply operators in a
separated form,~\cite{Beylkin:2002vla,Beylkin:2007gv} by rewriting the 
full multi-dimensional operator as a product of one-dimensional
contributions. For many important operators such a separation is not
exact, but Beylkin and coworkers have shown that it can be achieved to
any predefined precision as an expansion in Gaussian
functions.~\cite{Beylkin:2002vla,Fann:2004ub,Beylkin:2005wz}
To apply such operators in multiple dimensions, and simultaneously
retain the local adaptivity in the representation of functions,
it is essential to employ the non-standard form of
operators,~\cite{Gines:1998uw,Beylkin:2008im} which, in contrast
to the standard one, allows to decouple length scales.

These combined efforts (\ac{MW} representation of functions and
operators, separable operator representations, non-standard form of
operators) made the accurate application of several important convolution
operators efficient in three dimensions:
\begin{equation}
  	\label{eq:intconv}
  	g(r) = \hat{G}_\mu f(r) = \int G_\mu(r-r') f(r') dr' \, .
\end{equation}
Among such operators are the Poisson ($\mu = 0$) and the \ac{BSH}
($\mu^2 > 0$) kernels:
\begin{equation}
  	\label{eq:greenskernel}
	G_{\mu}(r-r') = \frac{e^{-\mu |r-r'|}}{4\pi |r-r|} \, .
\end{equation}

This mathematical framework was introduced to the computational
chemistry community in the mid 2000's by Harrison and
co-workers.~\cite{Harrison:2003cn,Harrison:2004vua,Yanai:2004tr,Yanai:2004vo}
They demonstrated that \acp{MW} could be employed to solve the \ac{KS}
equations in their integral reformulation:~\cite{Kalos:1962uj}
\begin{equation}\label{eq:harrisonw}
  	\orbital_i = -2 \hat{G}_{\mu_i} \potential \orbital_i \, ,
\end{equation}
where the $\orbital_i$'s are the \ac{KS} orbitals and the potential operator
$\potential$ includes external (nuclear), Hartree, exchange and correlation
contributions, while the kinetic operator and the orbital energies are
included in the \ac{BSH} operator as 
$2 \hat{G}_{\mu_i} = \big(\kinetic - \epsilon_i\big)^{-1}$, with 
$\mu_i^2 = -2\epsilon_i$. The ordinary \ac{KS} equations 
$\fockOper \orbital_i = \epsilon_i \orbital_i$, where 
$\fockOper = \kinetic + \potential$ is the \ac{KS} Hamiltonian,
can be recovered by recalling that 
$(\nabla^2-\mu^2) G_{\mu}(r-r') = -\delta(r-r')$. Such an integral formalism,
when combined with a Krylov subspace accelerator,~\cite{Harrison:2004gd}
leads to fast and robust convergence of the fix-point iteration of 
Eq.~(\ref{eq:harrisonw}).

The integral formulation, in combination with \acp{MW}, provides unprecedented
accuracy for all-electron calculations, without relying on molecular
symmetry,~\cite{Yanai:2004vo,Yanai:2004tr,Harrison:2004vua} and has also been
extended to excited states~\cite{Yanai:2005en,Yanai:2015gb,Kottmann:2015gc} and
electric~\cite{Sekino:2008ef,Sekino:2012ge} and magnetic~\cite{Jensen:2016jy}
linear response properties. In this approach all functions and operators, such as
orbitals, densities and potential energy contributions to the Kohn--Sham Hamiltonian,
are represented using Multiwavelets. Concerning potential energy terms, the external
potential is obtained by projection~\footnote{For the singularity of the
nuclear potential, a simple smoothing is employed, but its effect on the
accuracy is easily controlled by a single parameter~\cite{Harrison:2004vua}.}
onto the \ac{MW} basis, the Hartree potential is computed through Poisson's
equation:

\begin{equation}
	V_{\text{Hartree}}(r) = \int \frac{\rho(r')}{4\pi|r-r'|}dr' \, ,
\end{equation}
and the \ac{XC} potential is computed explicitly in the \ac{MW} representation
from the following expression:\footnote{Displayed as spin-unpolarized for clarity.
Extension to spin-\ac{DFT} is fairly straightforward.}
\begin{equation}
	\label{eq:xcpot}
	V_{\text{XC}}(r) = \frac{\partial f_{\text{XC}}}{\partial \rho} -
    2\nabla\frac{\partial f_{\text{XC}}}{\partial |\nabla\rho|^2}\cdot \nabla\rho \, .
\end{equation}

The partial derivatives of the \ac{XC} kernel $f_{XC}$ can be mapped point-wise
though external \ac{XC} libraries,~\cite{marques2012libxc,Ekstrom:2010bz} and
the gradients are computed by the approach of Alpert~\etal\cite{Alpert:1999tk}
For hybrid functionals, a fraction of the exact Hartree-Fock exchange contribution
is included in the \ac{KS} Hamiltonian (25\% for PBE0). In our work we follow the
method proposed by Yanai and coworkers,~\cite{Yanai:2004tr} where the exchange
operator is defined as:
\begin{equation}
  	\hat{K} f(r) = \sum_j \orbital_j(r) \int \frac{\orbital_j(r')f(r')}{4\pi|r-r'|}\ud r' \, .
\end{equation}
The above expression is again computed directly within the \ac{MW} framework
through repeated application of the Poisson operator to different orbital products.

While \acp{MW} are able to provide high-accuracy solution of integral equations in the
form of Eq. (\ref{eq:intconv}), the same is not true for differential operators. In particular,
high-order derivatives should be avoided in order to maintain accuracy in numerical algorithms.
For this reason, we have found that the direct evaluation of the kinetic energy as a 2nd
derivative of the wave function does not give the desired accuracy. Instead, we avoid the
kinetic operator by computing the update to the eigenvalue directly:\footnote{This update
is exact, provided that the orbital update comes directly from the application of
the \ac{BSH} operator defined by the previous (not necessarily exact) eigenvalue:
$\Delta\orbital^n = -2\hat{G}^n_\mu\left[\hat{V}^n\orbital^n\right] - \orbital^n$.
Generalizations can be made for multiple orbitals.}
\begin{equation}
  \Delta\epsilon^n
  = \frac{\langle\orbital^{n+1}|\hat{V}^n|\Delta\orbital^n\rangle}
  	{\langle\orbital^{n+1}|\orbital^{n+1}\rangle}
  + \frac{\langle\orbital^{n+1}|\Delta\hat{V}^n|\orbital^{n+1}\rangle}
  	{\langle\orbital^{n+1}|\orbital^{n+1}\rangle},
\end{equation}
where the $\Delta$'s refer to differences between iterations $n$ and $n+1$. In contrast,
the gradients in the expression for the \ac{GGA} potential in Eq.~(\ref{eq:xcpot}) have
not been found to affect the accuracy, partly because of a slightly conservative over
representation of the density grid,\footnote{The grid is constructed such that it holds
both the density and its gradient within the requested accuracy.} and partly because the
\ac{XC} energy is (by construction) only a small part of the total energy, thus reducing
its relative accuracy requirement.

\begin{figure}[t!]
\includegraphics[width=1.\columnwidth,angle=0]{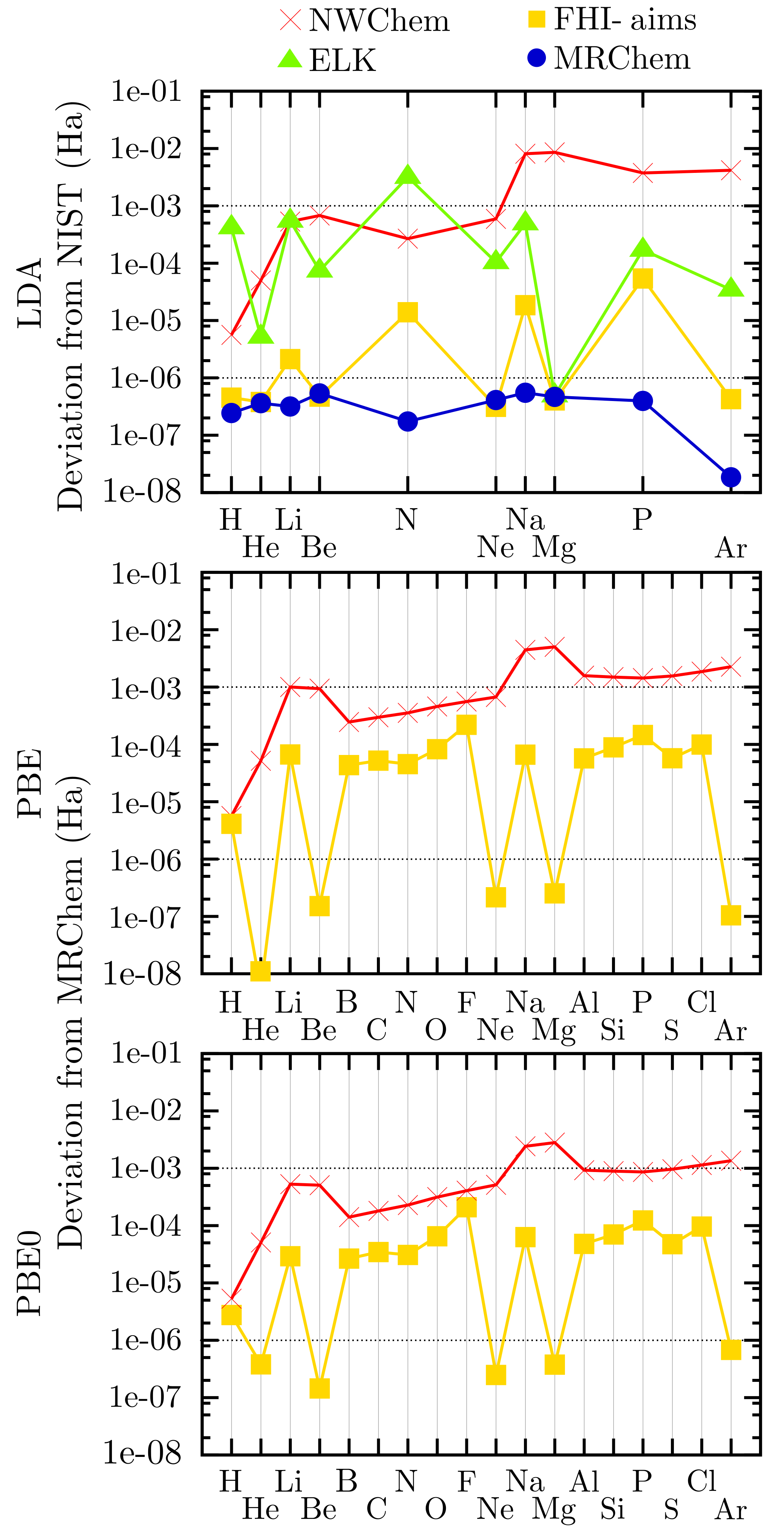}
\caption{Absolute deviations in total energy found for different functionals
	for selected atoms. For LDA-SVWN5, energy differences are w.r.t. NIST
    all-electron values.~\cite{kotochigova1997atomic} For GGA-PBE and hybrid-PBE0,
    the energy differences are w.r.t \mrchem{}. In all codes the largest basis
	set and tighter parameters were used. In all plots the reference values
	(NIST for LDA and \mrchem{} for PBE/PBE0) are given with 6 decimal precision;
	a displayed error below 1e-06 Ha means that no discrepancy is detectable.}
	\label{fig:single_energy}
\end{figure}

\begin{figure*}[t!]
\includegraphics[width=2.1\columnwidth,angle=0]{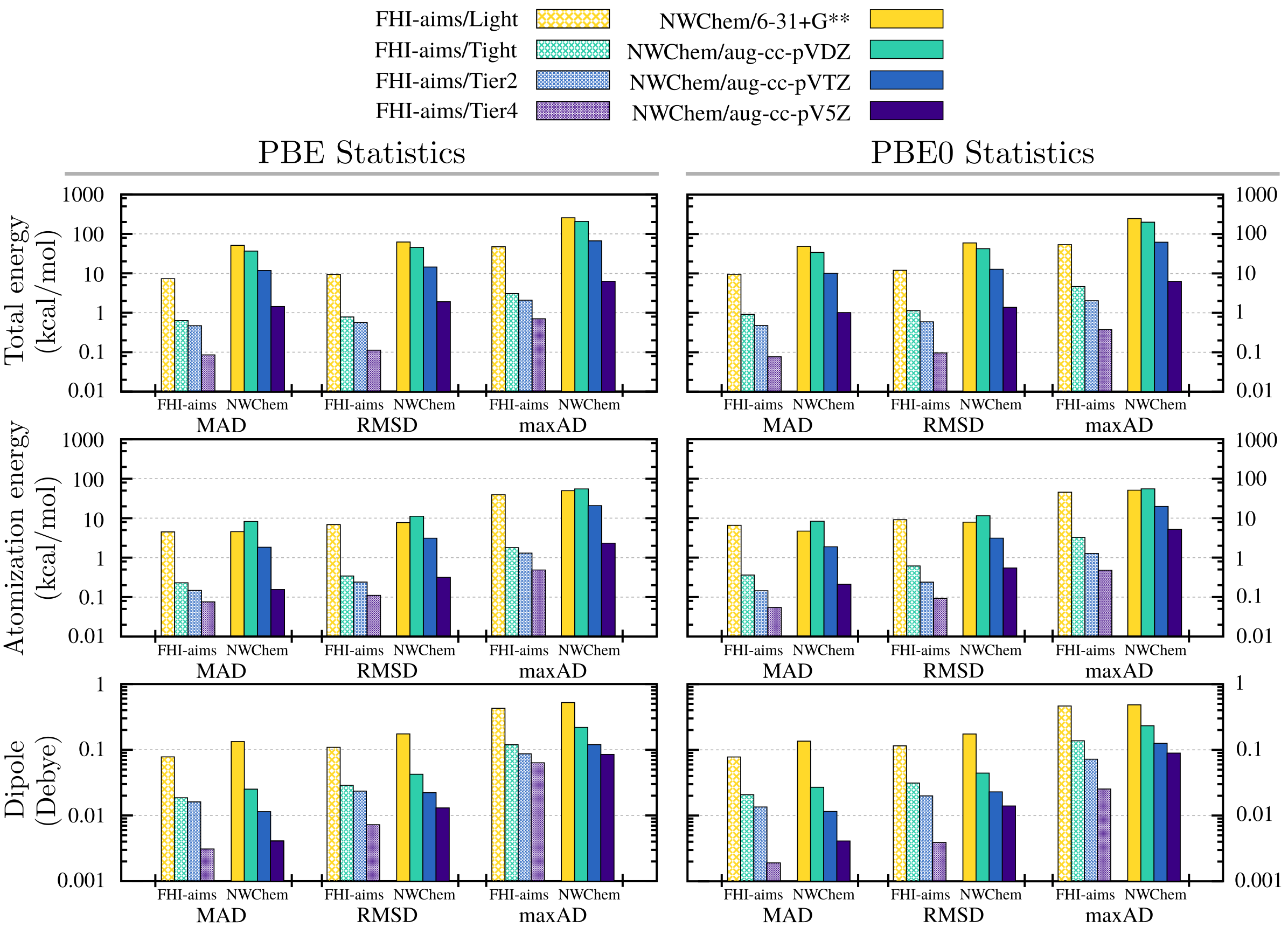}
\caption{GGA-PBE (left) and hybrid-PBE0 (right) deviations in total energy,
	atomization energy, and electrostatic dipole moment for the set of 210
	molecules with respect to highly accurate values obtained using \mrchem{}. 
	MAD, RMSD and maxAD stand for mean absolute deviation, root mean square 
	deviation and maximum absolute deviation, respectively. Results are included
	for two different DFT codes (\nwchem{} and \fhi{}) and four bases each
	(ranging from light/standard to the largest available).}
  	\label{fig:molecules-stats}
\end{figure*}

%\subsection{Computational details}\label{sec:comp-deta}

In this work, the \ac{MW} calculations are performed with \mrchem{},~\cite{mrchem}
the \ac{GTO}~\cite{feller1996role,schuchardt2007basis} calculations with
\nwchem{}~\cite{nwchem} and the \ac{NAO}~\cite{delley1990all} calculations
with \fhi{}.~\cite{fhiaims,1367-2630-14-5-053020} \ac{APW+lo} calculations
are performed with \elk{}.~\cite{elk} The exchange-correlation functionals are
calculated using the \libxc{}~\cite{marques2012libxc} library in case of \nwchem{}
and \fhi{}, and the \xcfun{}~\cite{Ekstrom:2010bz} library for \mrchem{}.

The raw data of our study, as well as instructions for its reproducibility
is available in the \ac{SI}.\cite{0EM0EL_2017}
Our test set comprises 211 molecules. In addition to the 147 systems 
from the G2/97 test set~\cite{schmider1998optimized} containing light
elements up to the third row, it contains molecules with chemical elements
that are underrepresented in the G2/97 test set (Be, Li, Mg, Al, F, Na,
S and Cl) as well as 6 non-bonded systems. For most of the systems, the
experimental structure obtained from the NIST Computational Chemistry
Comparison and Benchmark Database~\cite{NIST} was employed. In the
remaining cases, geometries have been optimized at the MP2 level of
theory, using the largest Gaussian basis set (see \ac{SI}\cite{0EM0EL_2017}
for details).

We have considered four different basis sets each within \nwchem{} and
\fhi{}, which include {\it small ones} intended for prerelaxations and
energy differences between bonded structures ("light", 6-31+G**),
{\it production} basis sets considered in most publications ("tight",
``tier2'' for \fhi{} and aug-cc-pVDZ, aug-cc-pVTZ for Gaussian codes),
as well the {\it largest available} basis set: ``tier4'' for \fhi{}
(``tier3'' for H) and aug-cc-pV5Z for \nwchem{}.\footnote{For Li, Be,
  Na and Mg the largest basis set is aug-cc-pVQZ and has therefore
  been employed. For the other elements, the corresponding 6Z basis is
  also available, but attempts to employ such a basis led to
  overcompleteness problems, often resulting in energies higher than
  the 5Z results.}

An accurate, global resolution-of-identity approach ("RI-V" in
Ref.\cite{1367-2630-14-5-053020}) is employed to evaluate the four-center
Coulomb operator in hybrid-PBE0 in \fhi{}. It is important to note that
the \ac{NAO} basis sets include a ``minimal basis'' of atomic radial
functions determined for the same \ac{XC} functional as used later in
the three-dimensional SCF calculations. This is standard practice in
\fhi{} for semilocal density functionals. For hybrid-PBE0, these radial
functions are provided by linking \fhi{} to the "atom\_sphere" atomic
solver code for spherically symmetric free atoms developed in the
Goedecker group for several years.\cite{transferability}

%\section{Results}

%\subsection{Total energy of isolated atoms}\label{sec:energy-isol-atoms}

The ground state energy of atoms from Hydrogen ($Z$=1) to Argon
($Z$=18) has been computed with the three chosen functionals. Our
results are summarized in Figure~\ref{fig:single_energy}.  For all
computational methods employed, the results of this section refer to
the most accurate basis set employed (see previous section and \ac{SI}\cite{0EM0EL_2017}
for details). The top panel reports the LDA-SVWN5 values as
absolute errors with respect to the reference values of the
NIST~\cite{kotochigova1997atomic} database for non-relativistic, spin
polarized, spherically symmetric atoms. As expected, \acp{MW} yield
differences which are consistently below the requested accuracy of 1
$\mu$\acs{Hartree}. The \ac{NAO} and \ac{APW+lo} approaches achieve
average errors of $\sim$0.01-0.1\,m\acs{Hartree} and
$\sim$0.1-1\,m\acs{Hartree}, respectively. \acp{GTO} are limited to
around m\acs{Hartree} accuracy. The \ac{GTO} outliers (Li, Be, Na and
Mg) have been computed with the aug-cc-pVQZ basis, because the
aug-cc-pV5Z basis set is not available for these elements. Had
5Z-quality functions been available for all elements, a more uniform
error for \acs{GTO} would have resulted along the series, but the
overall picture would only improve slightly.

In the GGA-PBE (middle) and hybrid-PBE0 (bottom) panels, all 18 atoms
(both spherical and not), are included. The non-relativistic,
spin-polarized electronic density and the total energy of the ground state,
computed using \mrchem{} (converged within $\mu$\ac{Hartree})
serves as the reference to which the other approaches are compared.
For both functionals \nwchem{} performs at the limit of chemical
accuracy ($\sim$ 1\,m\ac{Hartree}). The \acp{NAO} in \fhi{}
achieve 0.1\,m\ac{Hartree} or better, except for fluorine
(0.3\,m\ac{Hartree}).  For closed-shell atoms, \fhi{} is
essentially exact because the exact radial functions of spherically
symmetric, spin-unpolarized atoms are included in the basis sets.
For \acp{GTO}, we observe that the total energy error grows with the
atomic number, $Z$. In contrast, the accuracy of \acp{NAO} is less
affected by the nuclear charge, with errors generally below
0.1~m\ac{Hartree} for the $Z$ range examined here, irrespective of the
choice of functional. For \ac{APW+lo}, only the LDA-SVWN5 values are
included in Figure \ref{fig:single_energy}: the corresponding GGA-PBE
and hybrid-PBE0 errors achieved in this work are above the threshold
of 1e-03\,Ha (dashed line) and were not considered further because it
is unclear how much they might be affected by implementation-specific
aspects other than the basis set.

%\subsection{Molecular calculations}\label{sec:molec-calc}
 
The total energies, atomization energies and dipole moments of the 211
molecules considered have been computed within the GGA-PBE and hybrid-PBE0
functionals using \mrchem{} with the highest affordable precision
(below 1 $\mu$\ac{Hartree} throughout). Fig.~\ref{fig:molecules-stats}
reports the \ac{MAD}, \ac{RMSD} and \ac{maxAD} obtained for total energy
(top panel), atomization energy (medium panel) and dipole moment (bottom
panel) w.r.t \mrchem{} for the GGA-PBE and hybrid-PBE0 functionals,
respectively.\footnote{Due to technical reasons in convergence,
CH$_3$CH$_2$O was excluded from the PBE0 results, while CCH was
excluded in both PBE and PBE0.} For all the molecules, the correct
ground-state spin multiplicity was specified.

Total energies are a measure of the accuracy achieved by each
method/basis pair, whereas the atomization energies deserve special
attention for their role in the development of density functionals,
generally benchmarked against such thermodynamic values. However,
as recently pointed out by Medvedev~\etal\cite{Medvedev:2017dj}
the variational energy is not the optimal measure for the quality
of the calculated electronic density, which influences numerous other
observables. For this reason, we have included the dipole moment as a
non-variational quantity in our benchmarks (dipole errors are linear
in the density error, whereas energies are quadratic). Dipole moments
also serve as a verification that the different methods converge
to the same electronic state and not to a nearby metastable configuration.
Although the existence of multiple metastable SCF solutions in
Kohn-Sham DFT is well known, it is often not  detected by users of
electronic structure codes. The solution strategy, also employed in
the present paper, is to probe different spin initializations of each
molecule to identify the global minimum. In the present work, the
correct identification has been validated by ensuring consistency of
the dipole moment as well as the \ac{KS} eigenvalue spectra produced
by the three distinct electronic structure methods.

Several important conclusions can be drawn from the results obtained:

\begin{enumerate}

\item For total energies, \fhi{} is able to reach more accurate
	results than  \nwchem{} with \acp{GTO}, for basis sets of comparable
    size (e.g. "tier4" vs. aug-cc-pV5Z).
    
\item For atomization energies, both \acp{NAO} and \acp{GTO} benefit from
	error cancellation to some extent. Such a cancellation is however much
    stronger for \acp{GTO} where the \ac{RMSD} is lowered by a factor 4-8
    in most cases, whereas for \acp{NAO} only by a factor of 1,5-2. In both
    cases the cancellation is more marked for the smallest bases. Despite
    the smaller cancellations, \acp{NAO} are still closer to the converged
    limit than \acp{GTO}, for comparable basis sets.
    
\item The two functionals considered (GGA-PBE and hybrid-PBE0) yield
  	very similar results, and we therefore assume that our conclusions
    concerning the accuracy of the different approaches (\acp{NAO} and
    \acp{GTO}) will hold also for other functionals of the same type.
    
\item Dipole moments can be considered accurate if deviations are below
	0.01\,Debye.~\cite{bak2000accuracy,NIST} Only the largest basis sets
    in \ac{NAO} and \ac{GTO} used in our calculations achieve this target
    on average, but even such basis sets have outliers with errors close
    to 0.1\,Debye.
    
\item Due to the cumbersome convergence of periodic \ac{DFT} codes with
	respect to the box size, we did not include \ac{APW+lo} results for
    the entire test set of molecules. Nevertheless, for a small subset
    of molecules for which the limit of the box size was reached, we
    found atomization energies with errors of about 1\,kcal/mol (see
    \ac{SI}\cite{0EM0EL_2017}). Our experience suggests that it is
    technically challenging for \ac{APW}-based codes to reach accuracies
    below 1\,kcal/mol on atomization energies.
    
\end{enumerate}

As a final remark, we stress that for a few atoms (Li, Be, Na, Mg),
the aug-cc-pV5Z basis is not available, as previously mentioned in the
atomic calculations part. Had it been available, \acp{GTO} might have
yielded somewhat higher precision for the affected systems than in
our benchmarks. However, considering the large size of our sample, the
fact that only a few atoms in a molecule are affected, and the small
improvement that can be inferred from the atomic calculations, our
main conclusions still hold. On the other hand, such a \emph{de facto}
limitation of the availability of \ac{GTO} basis sets illustrates how
demanding it is to generate such basis sets. In contrast, \acp{MW} and
\acp{NAO} are much less affected by such a limitation.

%\section{Conclusions}

To the best of our knowledge, this work presents the most accurate atomization
energies calculated to date, for a large benchmark set of molecules. We 
conclude that moderately sized \ac{GTO} basis sets, frequently used in
quantum chemistry applications, suffer from average total energy errors
much larger than 10~kcal/mol, and while very large \ac{GTO} basis sets
yield the desired accuracy on average, there are still significant
outliers. Furthermore, it may not always be feasible to employ such
basis sets for systems much larger than those included in this study. 

\acp{NAO} give much better accuracy even for moderately large bases
(``tight'' and beyond) since they can be constructed to possess the
numerically correct behavior for a given XC functional, both in the nuclear
as well as in the tail region. When feasible, \ac{APW+lo}-based calculations
achieve errors around 1~m\acs{Hartree} for total energies, and 1~kcal/mol
for atomization energies. However, this level of convergence is difficult
to reach for general molecular systems.

Our results show that the basis set error can dominate
over errors arising from the choice of \ac{XC} functional under many
circumstances, in particular if some of the most advanced and accurate
functionals are used. Our results set new standards in the verification
and validation of electronic structure methods, and we expect them to be
used to assess the accuracy of all future developments in \acl{DFT} methods.

\begin{acknowledgments}
This research was partly supported by the Research Council of Norway through
a Centre of Excellence Grant (Grant No. 179568/V30) with computing time
provided by NOTUR (Grant No. NN4654K), and partly by the NCCR MARVEL, funded
by the Swiss National Science Foundation (SNSF) with computing time provided
by CSCS under project s707. S.~Saha acknowledges support from the SNSF. J.A.F.-L.
acknowledges fruitful discussions with John Kay Dewhurst on the APW method.
\end{acknowledgments}

\bibliographystyle{apsrev4-1}
\bibliography{manuscript_jpcl}

\begin{acronym}
\acro{Hartree}[Ha]{Hartree}
\acro{CBS}{complete basis set}
\acro{MRA}{Multiresolution Analysis}
\acro{MW}{Multiwavelet}
\acro{SI}{Supporting Information}
\acro{GTO}{Gaussian-type orbital}
\acro{BSH}{Bound-State Helmholtz}
\acro{SCF}{Self-Consistent Field}
\acro{HF}{Hartree--Fock}  
\acro{KS}{Kohn--Sham}
\acro{XC}{Exchange-Correlation}
\acro{DFT}{Density Functional Theory}  
\acro{GGA}{Generalized Gradient Approximation}
\acro{LDA}{Local Density Approximation}
\acro{KAIN}{Krylov-Accelerated Inexact Newton}
\acro{AO}{Atomic Orbital}
\acro{MO}{Molecular Orbital}
\acro{LCAO}{Linear Combination of Atomic Orbitals}
\acro{NAO}{numeric atom-centered orbital}
\acro{LAPW}{linearised augmented plane wave}
\acro{APW}{augmented plane wave}
\acro{APW+lo}{APW+local orbital}
\acro{MAD}{Mean Absolute Deviation}
\acro{MD}{Mean Deviation}
\acro{maxAD}{Maximum Absolute Deviation}
\acro{minAD}{Minimum Absolute Deviation}
\acro{RMSD}{Root Mean Square Deviation}
\acro{SD}{Standard Deviation}
\acro{MBPT}{Many-body perturbation theory}
\acro{MP2}{2nd-order M\o ller-Plesset}
\acro{CC}{Coupled Cluster}
\acro{CI}{Configuration Interaction}
\acro{CCSD}{Coupled-Cluster Singles Doubles}
\acro{MCSCF}{Multi-Configurational Self-Consistent Field}
\end{acronym}

\end{document}

%% file: newcommands.tex
\newcommand{\etal}{{\emph{et al.}\ }}

\newlength{\dch}
\newlength{\sch}
\newcommand{\ud}{\ensuremath{\,\mathrm{d}}}

\newcommand{\scaling}[2]{\ensuremath{\phi^{#1}_{#2}}}

\newcommand{\orbital}{\ensuremath{\varphi}}

\newcommand{\potential}{\ensuremath{\hat{V}}}
\newcommand{\kinetic}{\ensuremath{\hat{T}}}

\newcommand{\fockOper}{\ensuremath{\hat{H}}}

%% file: manuscript_jpcl.bbl
\begin{thebibliography}{74}%
\makeatletter
\providecommand \@ifxundefined [1]{%
 \@ifx{#1\undefined}
}%
\providecommand \@ifnum [1]{%
 \ifnum #1\expandafter \@firstoftwo
 \else \expandafter \@secondoftwo
 \fi
}%
\providecommand \@ifx [1]{%
 \ifx #1\expandafter \@firstoftwo
 \else \expandafter \@secondoftwo
 \fi
}%
\providecommand \natexlab [1]{#1}%
\providecommand \enquote  [1]{``#1''}%
\providecommand \bibnamefont  [1]{#1}%
\providecommand \bibfnamefont [1]{#1}%
\providecommand \citenamefont [1]{#1}%
\providecommand \href@noop [0]{\@secondoftwo}%
\providecommand \href [0]{\begingroup \@sanitize@url \@href}%
\providecommand \@href[1]{\@@startlink{#1}\@@href}%
\providecommand \@@href[1]{\endgroup#1\@@endlink}%
\providecommand \@sanitize@url [0]{\catcode `\\12\catcode `\$12\catcode
  `\&12\catcode `\#12\catcode `\^12\catcode `\_12\catcode `\%12\relax}%
\providecommand \@@startlink[1]{}%
\providecommand \@@endlink[0]{}%
\providecommand \url  [0]{\begingroup\@sanitize@url \@url }%
\providecommand \@url [1]{\endgroup\@href {#1}{\urlprefix }}%
\providecommand \urlprefix  [0]{URL }%
\providecommand \Eprint [0]{\href }%
\providecommand \doibase [0]{http://dx.doi.org/}%
\providecommand \selectlanguage [0]{\@gobble}%
\providecommand \bibinfo  [0]{\@secondoftwo}%
\providecommand \bibfield  [0]{\@secondoftwo}%
\providecommand \translation [1]{[#1]}%
\providecommand \BibitemOpen [0]{}%
\providecommand \bibitemStop [0]{}%
\providecommand \bibitemNoStop [0]{.\EOS\space}%
\providecommand \EOS [0]{\spacefactor3000\relax}%
\providecommand \BibitemShut  [1]{\csname bibitem#1\endcsname}%
\let\auto@bib@innerbib\@empty
%</preamble>
\bibitem [{\citenamefont {Hohenberg}\ and\ \citenamefont
  {Kohn}(1964)}]{hohenberg-kohn}%
  \BibitemOpen
  \bibfield  {author} {\bibinfo {author} {\bibfnamefont {H.}~\bibnamefont
  {Hohenberg}}\ and\ \bibinfo {author} {\bibfnamefont {W.}~\bibnamefont
  {Kohn}},\ }\href@noop {} {\bibfield  {journal} {\bibinfo  {journal} {Phys.
  Rev.}\ }\textbf {\bibinfo {volume} {136}},\ \bibinfo {pages} {B864} (\bibinfo
  {year} {1964})}\BibitemShut {NoStop}%
\bibitem [{\citenamefont {Kohn}\ and\ \citenamefont {Sham}(1965)}]{kohn-sham}%
  \BibitemOpen
  \bibfield  {author} {\bibinfo {author} {\bibfnamefont {W.}~\bibnamefont
  {Kohn}}\ and\ \bibinfo {author} {\bibfnamefont {L.~J.}\ \bibnamefont
  {Sham}},\ }\href {\doibase 10.1103/PhysRev.140.A1133} {\bibfield  {journal}
  {\bibinfo  {journal} {Phys. Rev.}\ }\textbf {\bibinfo {volume} {140}},\
  \bibinfo {pages} {A1133} (\bibinfo {year} {1965})}\BibitemShut {NoStop}%
\bibitem [{\citenamefont {Goerigk}\ and\ \citenamefont
  {Grimme}(2010)}]{Goerigk:2010kc}%
  \BibitemOpen
  \bibfield  {author} {\bibinfo {author} {\bibfnamefont {L.}~\bibnamefont
  {Goerigk}}\ and\ \bibinfo {author} {\bibfnamefont {S.}~\bibnamefont
  {Grimme}},\ }\href {\doibase 10.1021/ct900489g} {\bibfield  {journal}
  {\bibinfo  {journal} {J. Chem. Theory Comput.}\ }\textbf {\bibinfo {volume}
  {6}},\ \bibinfo {pages} {107} (\bibinfo {year} {2010})}\BibitemShut {NoStop}%
\bibitem [{\citenamefont {Karton}\ \emph {et~al.}(2008)\citenamefont {Karton},
  \citenamefont {Tarnopolsky}, \citenamefont {Lamere}, \citenamefont {Schatz},\
  and\ \citenamefont {Martin}}]{Karton:2008ho}%
  \BibitemOpen
  \bibfield  {author} {\bibinfo {author} {\bibfnamefont {A.}~\bibnamefont
  {Karton}}, \bibinfo {author} {\bibfnamefont {A.}~\bibnamefont {Tarnopolsky}},
  \bibinfo {author} {\bibfnamefont {J.-F.}\ \bibnamefont {Lamere}}, \bibinfo
  {author} {\bibfnamefont {G.~C.}\ \bibnamefont {Schatz}}, \ and\ \bibinfo
  {author} {\bibfnamefont {J.~M.~L.}\ \bibnamefont {Martin}},\ }\href {\doibase
  10.1021/jp801805p} {\bibfield  {journal} {\bibinfo  {journal} {J. Phys. Chem.
  A}\ }\textbf {\bibinfo {volume} {112}},\ \bibinfo {pages} {12868} (\bibinfo
  {year} {2008})}\BibitemShut {NoStop}%
\bibitem [{\citenamefont {Zhao}\ and\ \citenamefont
  {Truhlar}(2005)}]{Zhao:2005kj}%
  \BibitemOpen
  \bibfield  {author} {\bibinfo {author} {\bibfnamefont {Y.}~\bibnamefont
  {Zhao}}\ and\ \bibinfo {author} {\bibfnamefont {D.~G.}\ \bibnamefont
  {Truhlar}},\ }\href {\doibase 10.1021/jp050536c} {\bibfield  {journal}
  {\bibinfo  {journal} {J. Phys. Chem. A}\ }\textbf {\bibinfo {volume} {109}},\
  \bibinfo {pages} {5656} (\bibinfo {year} {2005})}\BibitemShut {NoStop}%
\bibitem [{\citenamefont {Perdew}\ \emph {et~al.}(1996)\citenamefont {Perdew},
  \citenamefont {Burke},\ and\ \citenamefont {Ernzerhof}}]{burke}%
  \BibitemOpen
  \bibfield  {author} {\bibinfo {author} {\bibfnamefont {J.}~\bibnamefont
  {Perdew}}, \bibinfo {author} {\bibfnamefont {K.}~\bibnamefont {Burke}}, \
  and\ \bibinfo {author} {\bibfnamefont {M.}~\bibnamefont {Ernzerhof}},\
  }\href@noop {} {\bibfield  {journal} {\bibinfo  {journal} {Phys. Rev. Lett.}\
  }\textbf {\bibinfo {volume} {77}},\ \bibinfo {pages} {3865} (\bibinfo {year}
  {1996})}\BibitemShut {NoStop}%
\bibitem [{\citenamefont {Perdew}\ and\ \citenamefont
  {Schmidt}(2001)}]{Perdew:2001}%
  \BibitemOpen
  \bibfield  {author} {\bibinfo {author} {\bibfnamefont {J.~P.}\ \bibnamefont
  {Perdew}}\ and\ \bibinfo {author} {\bibfnamefont {K.}~\bibnamefont
  {Schmidt}},\ }\href {\doibase http://dx.doi.org/10.1063/1.1390175} {\bibfield
   {journal} {\bibinfo  {journal} {AIP Conf. Proc.}\ }\textbf {\bibinfo
  {volume} {577}},\ \bibinfo {pages} {1} (\bibinfo {year} {2001})}\BibitemShut
  {NoStop}%
\bibitem [{\citenamefont {Li}\ \emph {et~al.}(2015)\citenamefont {Li},
  \citenamefont {Zheng}, \citenamefont {Cohen}, \citenamefont
  {Mori-S\'anchez},\ and\ \citenamefont {Yang}}]{PhysRevLett.114.053001}%
  \BibitemOpen
  \bibfield  {author} {\bibinfo {author} {\bibfnamefont {C.}~\bibnamefont
  {Li}}, \bibinfo {author} {\bibfnamefont {X.}~\bibnamefont {Zheng}}, \bibinfo
  {author} {\bibfnamefont {A.~J.}\ \bibnamefont {Cohen}}, \bibinfo {author}
  {\bibfnamefont {P.}~\bibnamefont {Mori-S\'anchez}}, \ and\ \bibinfo {author}
  {\bibfnamefont {W.}~\bibnamefont {Yang}},\ }\href {\doibase
  10.1103/PhysRevLett.114.053001} {\bibfield  {journal} {\bibinfo  {journal}
  {Phys. Rev. Lett.}\ }\textbf {\bibinfo {volume} {114}},\ \bibinfo {pages}
  {053001} (\bibinfo {year} {2015})}\BibitemShut {NoStop}%
\bibitem [{\citenamefont {Sun}\ \emph {et~al.}(2015)\citenamefont {Sun},
  \citenamefont {Ruzsinszky},\ and\ \citenamefont {Perdew}}]{sun2015strongly}%
  \BibitemOpen
  \bibfield  {author} {\bibinfo {author} {\bibfnamefont {J.}~\bibnamefont
  {Sun}}, \bibinfo {author} {\bibfnamefont {A.}~\bibnamefont {Ruzsinszky}}, \
  and\ \bibinfo {author} {\bibfnamefont {J.~P.}\ \bibnamefont {Perdew}},\
  }\href@noop {} {\bibfield  {journal} {\bibinfo  {journal} {Phys. Rev. Lett.}\
  }\textbf {\bibinfo {volume} {115}},\ \bibinfo {pages} {036402} (\bibinfo
  {year} {2015})}\BibitemShut {NoStop}%
\bibitem [{\citenamefont {Su}\ and\ \citenamefont {Xu}(2015)}]{QUA:QUA24849}%
  \BibitemOpen
  \bibfield  {author} {\bibinfo {author} {\bibfnamefont {N.~Q.}\ \bibnamefont
  {Su}}\ and\ \bibinfo {author} {\bibfnamefont {X.}~\bibnamefont {Xu}},\ }\href
  {\doibase 10.1002/qua.24849} {\bibfield  {journal} {\bibinfo  {journal} {Int.
  J. Quantum Chem.}\ }\textbf {\bibinfo {volume} {115}},\ \bibinfo {pages}
  {589} (\bibinfo {year} {2015})}\BibitemShut {NoStop}%
\bibitem [{\citenamefont {Peverati}\ and\ \citenamefont
  {Truhlar}(2014)}]{peverati2014quest}%
  \BibitemOpen
  \bibfield  {author} {\bibinfo {author} {\bibfnamefont {R.}~\bibnamefont
  {Peverati}}\ and\ \bibinfo {author} {\bibfnamefont {D.~G.}\ \bibnamefont
  {Truhlar}},\ }\href@noop {} {\bibfield  {journal} {\bibinfo  {journal}
  {Philos. Trans. R. Soc. London A: Math., Phys. and Eng. Sci.}\ }\textbf
  {\bibinfo {volume} {372}},\ \bibinfo {pages} {20120476} (\bibinfo {year}
  {2014})}\BibitemShut {NoStop}%
\bibitem [{\citenamefont {Mardirossian}\ and\ \citenamefont
  {Head-Gordon}(2016)}]{Head-Gordon_2016}%
  \BibitemOpen
  \bibfield  {author} {\bibinfo {author} {\bibfnamefont {N.}~\bibnamefont
  {Mardirossian}}\ and\ \bibinfo {author} {\bibfnamefont {M.}~\bibnamefont
  {Head-Gordon}},\ }\href {\doibase 10.1021/acs.jctc.6b00637} {\bibfield
  {journal} {\bibinfo  {journal} {J. Chem. Theory Comput.}\ }\textbf {\bibinfo
  {volume} {12}},\ \bibinfo {pages} {4303} (\bibinfo {year} {2016})},\ \bibinfo
  {note} {pMID: 27537680},\ \Eprint
  {http://arxiv.org/abs/http://dx.doi.org/10.1021/acs.jctc.6b00637}
  {http://dx.doi.org/10.1021/acs.jctc.6b00637} \BibitemShut {NoStop}%
\bibitem [{\citenamefont {Singh}\ and\ \citenamefont
  {Nordstrom}(2006)}]{singh2006planewaves}%
  \BibitemOpen
  \bibfield  {author} {\bibinfo {author} {\bibfnamefont {D.~J.}\ \bibnamefont
  {Singh}}\ and\ \bibinfo {author} {\bibfnamefont {L.}~\bibnamefont
  {Nordstrom}},\ }\href@noop {} {\emph {\bibinfo {title} {Planewaves,
  Pseudopotentials, and the LAPW method}}}\ (\bibinfo  {publisher} {Springer
  Science \& Business Media},\ \bibinfo {year} {2006})\BibitemShut {NoStop}%
\bibitem [{\citenamefont {Goedecker}\ and\ \citenamefont
  {Maschke}(1992)}]{transferability}%
  \BibitemOpen
  \bibfield  {author} {\bibinfo {author} {\bibfnamefont {S.}~\bibnamefont
  {Goedecker}}\ and\ \bibinfo {author} {\bibfnamefont {K.}~\bibnamefont
  {Maschke}},\ }\href@noop {} {\bibfield  {journal} {\bibinfo  {journal} {Phy.
  Rev. A}\ }\textbf {\bibinfo {volume} {45}},\ \bibinfo {pages} {88} (\bibinfo
  {year} {1992})}\BibitemShut {NoStop}%
\bibitem [{\citenamefont {Willand}\ \emph {et~al.}(2013)\citenamefont
  {Willand}, \citenamefont {Kvashnin}, \citenamefont {Genovese}, \citenamefont
  {Mayagoitia}, \citenamefont {Deb}, \citenamefont {Sadeghi}, \citenamefont
  {Deutsch},\ and\ \citenamefont {Goedecker}}]{NLCC}%
  \BibitemOpen
  \bibfield  {author} {\bibinfo {author} {\bibfnamefont {A.}~\bibnamefont
  {Willand}}, \bibinfo {author} {\bibfnamefont {Y.~O.}\ \bibnamefont
  {Kvashnin}}, \bibinfo {author} {\bibfnamefont {L.}~\bibnamefont {Genovese}},
  \bibinfo {author} {\bibfnamefont {{\' A}.~V.}\ \bibnamefont {Mayagoitia}},
  \bibinfo {author} {\bibfnamefont {A.~K.}\ \bibnamefont {Deb}}, \bibinfo
  {author} {\bibfnamefont {A.}~\bibnamefont {Sadeghi}}, \bibinfo {author}
  {\bibfnamefont {T.}~\bibnamefont {Deutsch}}, \ and\ \bibinfo {author}
  {\bibfnamefont {S.}~\bibnamefont {Goedecker}},\ }\href@noop {} {\bibfield
  {journal} {\bibinfo  {journal} {J. Chem. Phys.}\ }\textbf {\bibinfo {volume}
  {138}},\ \bibinfo {pages} {104109} (\bibinfo {year} {2013})}\BibitemShut
  {NoStop}%
\bibitem [{\citenamefont {Lejaeghere}\ \emph {et~al.}(2016)\citenamefont
  {Lejaeghere}, \citenamefont {Bihlmayer}, \citenamefont {Bj{\"o}rkman},
  \citenamefont {Blaha}, \citenamefont {Bl{\"u}gel}, \citenamefont {Blum},
  \citenamefont {Caliste}, \citenamefont {Castelli}, \citenamefont {Clark},
  \citenamefont {Dal~Corso} \emph {et~al.}}]{lejaeghere2016reproducibility}%
  \BibitemOpen
  \bibfield  {author} {\bibinfo {author} {\bibfnamefont {K.}~\bibnamefont
  {Lejaeghere}}, \bibinfo {author} {\bibfnamefont {G.}~\bibnamefont
  {Bihlmayer}}, \bibinfo {author} {\bibfnamefont {T.}~\bibnamefont
  {Bj{\"o}rkman}}, \bibinfo {author} {\bibfnamefont {P.}~\bibnamefont {Blaha}},
  \bibinfo {author} {\bibfnamefont {S.}~\bibnamefont {Bl{\"u}gel}}, \bibinfo
  {author} {\bibfnamefont {V.}~\bibnamefont {Blum}}, \bibinfo {author}
  {\bibfnamefont {D.}~\bibnamefont {Caliste}}, \bibinfo {author} {\bibfnamefont
  {I.~E.}\ \bibnamefont {Castelli}}, \bibinfo {author} {\bibfnamefont {S.~J.}\
  \bibnamefont {Clark}}, \bibinfo {author} {\bibfnamefont {A.}~\bibnamefont
  {Dal~Corso}},  \emph {et~al.},\ }\href@noop {} {\bibfield  {journal}
  {\bibinfo  {journal} {Science}\ }\textbf {\bibinfo {volume} {351}},\ \bibinfo
  {pages} {aad3000} (\bibinfo {year} {2016})}\BibitemShut {NoStop}%
\bibitem [{\citenamefont {Moncrieff}\ and\ \citenamefont
  {Wilson}(2005)}]{Moncrieff:2005tf}%
  \BibitemOpen
  \bibfield  {author} {\bibinfo {author} {\bibfnamefont {D.}~\bibnamefont
  {Moncrieff}}\ and\ \bibinfo {author} {\bibfnamefont {S.}~\bibnamefont
  {Wilson}},\ }\href
  {http://onlinelibrary.wiley.com/doi/10.1002/qua.20275/full} {\bibfield
  {journal} {\bibinfo  {journal} {Int. J. Quantum Chem.}\ }\textbf {\bibinfo
  {volume} {101}},\ \bibinfo {pages} {363} (\bibinfo {year}
  {2005})}\BibitemShut {NoStop}%
\bibitem [{\citenamefont {Kotochigova}\ \emph {et~al.}(1997)\citenamefont
  {Kotochigova}, \citenamefont {Levine}, \citenamefont {Shirley}, \citenamefont
  {Stiles},\ and\ \citenamefont {Clark}}]{kotochigova1997atomic}%
  \BibitemOpen
  \bibfield  {author} {\bibinfo {author} {\bibfnamefont {S.}~\bibnamefont
  {Kotochigova}}, \bibinfo {author} {\bibfnamefont {Z.~H.}\ \bibnamefont
  {Levine}}, \bibinfo {author} {\bibfnamefont {E.~L.}\ \bibnamefont {Shirley}},
  \bibinfo {author} {\bibfnamefont {M.~D.}\ \bibnamefont {Stiles}}, \ and\
  \bibinfo {author} {\bibfnamefont {C.~W.}\ \bibnamefont {Clark}},\ }\href
  {http://www.nist.gov/pml/data/dftdat/ptable.cfm} {\emph {\bibinfo {title}
  {Atomic reference data for electronic structure calculations}}}\ (\bibinfo
  {publisher} {National Institute of Standards and Technology, Physics
  Laboratory},\ \bibinfo {year} {1997})\BibitemShut {NoStop}%
\bibitem [{\citenamefont {Ballone}\ and\ \citenamefont
  {Galli}(1990)}]{Ballone:1990}%
  \BibitemOpen
  \bibfield  {author} {\bibinfo {author} {\bibfnamefont {P.}~\bibnamefont
  {Ballone}}\ and\ \bibinfo {author} {\bibfnamefont {G.}~\bibnamefont
  {Galli}},\ }\href {\doibase 10.1103/PhysRevB.42.1112} {\bibfield  {journal}
  {\bibinfo  {journal} {Phys. Rev. B}\ }\textbf {\bibinfo {volume} {42}},\
  \bibinfo {pages} {1112} (\bibinfo {year} {1990})}\BibitemShut {NoStop}%
\bibitem [{\citenamefont {Becke}(1992)}]{Becke:1992}%
  \BibitemOpen
  \bibfield  {author} {\bibinfo {author} {\bibfnamefont {A.~D.}\ \bibnamefont
  {Becke}},\ }\href {\doibase http://dx.doi.org/10.1063/1.462066} {\bibfield
  {journal} {\bibinfo  {journal} {J. Chem. Phys.}\ }\textbf {\bibinfo {volume}
  {96}},\ \bibinfo {pages} {2155} (\bibinfo {year} {1992})}\BibitemShut
  {NoStop}%
\bibitem [{\citenamefont {Daykov}\ \emph {et~al.}(2003)\citenamefont {Daykov},
  \citenamefont {Arias},\ and\ \citenamefont {Engeness}}]{arias}%
  \BibitemOpen
  \bibfield  {author} {\bibinfo {author} {\bibfnamefont {I.~P.}\ \bibnamefont
  {Daykov}}, \bibinfo {author} {\bibfnamefont {T.~A.}\ \bibnamefont {Arias}}, \
  and\ \bibinfo {author} {\bibfnamefont {T.~D.}\ \bibnamefont {Engeness}},\
  }\href {\doibase {10.1103/PhysRevLett.90.216402}} {\bibfield  {journal}
  {\bibinfo  {journal} {{Phys. Rev. Lett.}}\ }\textbf {\bibinfo {volume}
  {{90}}} (\bibinfo {year} {{2003}}),\
  {10.1103/PhysRevLett.90.216402}}\BibitemShut {NoStop}%
\bibitem [{NIS(2015)}]{NIST}%
  \BibitemOpen
  \href@noop {} {\enquote {\bibinfo {title} {{NIST Computational Chemistry
  Comparison and Benchmark Database, NIST Standard Reference Database Number
  101, Editor: Russell D. Johnson III}},}\ }\bibinfo {howpublished} {See
  http://cccbdb.nist.gov/} (\bibinfo {year} {2015})\BibitemShut {NoStop}%
\bibitem [{\citenamefont {Papas}\ and\ \citenamefont
  {Schaefer}(2006)}]{papas2006concerning}%
  \BibitemOpen
  \bibfield  {author} {\bibinfo {author} {\bibfnamefont {B.~N.}\ \bibnamefont
  {Papas}}\ and\ \bibinfo {author} {\bibfnamefont {H.~F.}\ \bibnamefont
  {Schaefer}},\ }\href@noop {} {\bibfield  {journal} {\bibinfo  {journal} {J.
  Mol. Struct.: THEOCHEM}\ }\textbf {\bibinfo {volume} {768}},\ \bibinfo
  {pages} {175} (\bibinfo {year} {2006})}\BibitemShut {NoStop}%
\bibitem [{\citenamefont {Harrison}\ \emph {et~al.}(2003)\citenamefont
  {Harrison}, \citenamefont {Fann}, \citenamefont {Yanai},\ and\ \citenamefont
  {Beylkin}}]{Harrison:2003cn}%
  \BibitemOpen
  \bibfield  {author} {\bibinfo {author} {\bibfnamefont {R.~J.}\ \bibnamefont
  {Harrison}}, \bibinfo {author} {\bibfnamefont {G.~I.}\ \bibnamefont {Fann}},
  \bibinfo {author} {\bibfnamefont {T.}~\bibnamefont {Yanai}}, \ and\ \bibinfo
  {author} {\bibfnamefont {G.}~\bibnamefont {Beylkin}},\ }\href {\doibase
  10.1007/3-540-44864-0_11} {\bibfield  {journal} {\bibinfo  {journal} {Comput.
  Sci. - ICCS 2003: Int. Conf. Proc.}\ ,\ \bibinfo {pages} {103}} (\bibinfo
  {year} {2003})}\BibitemShut {NoStop}%
\bibitem [{\citenamefont {Harrison}\ \emph {et~al.}(2004)\citenamefont
  {Harrison}, \citenamefont {Fann}, \citenamefont {Yanai}, \citenamefont
  {Gan},\ and\ \citenamefont {Beylkin}}]{Harrison:2004vua}%
  \BibitemOpen
  \bibfield  {author} {\bibinfo {author} {\bibfnamefont {R.~J.}\ \bibnamefont
  {Harrison}}, \bibinfo {author} {\bibfnamefont {G.~I.}\ \bibnamefont {Fann}},
  \bibinfo {author} {\bibfnamefont {T.}~\bibnamefont {Yanai}}, \bibinfo
  {author} {\bibfnamefont {Z.}~\bibnamefont {Gan}}, \ and\ \bibinfo {author}
  {\bibfnamefont {G.}~\bibnamefont {Beylkin}},\ }\href
  {http://eutils.ncbi.nlm.nih.gov/entrez/eutils/elink.fcgi?dbfrom=pubmed&id=15634124&retmode=ref&cmd=prlinks}
  {\bibfield  {journal} {\bibinfo  {journal} {J. Chem. Phys.}\ }\textbf
  {\bibinfo {volume} {121}},\ \bibinfo {pages} {11587} (\bibinfo {year}
  {2004})}\BibitemShut {NoStop}%
\bibitem [{\citenamefont {Yanai}\ \emph
  {et~al.}(2004{\natexlab{a}})\citenamefont {Yanai}, \citenamefont {Fann},
  \citenamefont {Gan}, \citenamefont {Harrison},\ and\ \citenamefont
  {Beylkin}}]{Yanai:2004tr}%
  \BibitemOpen
  \bibfield  {author} {\bibinfo {author} {\bibfnamefont {T.}~\bibnamefont
  {Yanai}}, \bibinfo {author} {\bibfnamefont {G.~I.}\ \bibnamefont {Fann}},
  \bibinfo {author} {\bibfnamefont {Z.}~\bibnamefont {Gan}}, \bibinfo {author}
  {\bibfnamefont {R.~J.}\ \bibnamefont {Harrison}}, \ and\ \bibinfo {author}
  {\bibfnamefont {G.}~\bibnamefont {Beylkin}},\ }\href
  {http://eutils.ncbi.nlm.nih.gov/entrez/eutils/elink.fcgi?dbfrom=pubmed&id=15473723&retmode=ref&cmd=prlinks}
  {\bibfield  {journal} {\bibinfo  {journal} {J. Chem. Phys.}\ }\textbf
  {\bibinfo {volume} {121}},\ \bibinfo {pages} {6680} (\bibinfo {year}
  {2004}{\natexlab{a}})}\BibitemShut {NoStop}%
\bibitem [{\citenamefont {Yanai}\ \emph
  {et~al.}(2004{\natexlab{b}})\citenamefont {Yanai}, \citenamefont {Fann},
  \citenamefont {Gan}, \citenamefont {Harrison},\ and\ \citenamefont
  {Beylkin}}]{Yanai:2004vo}%
  \BibitemOpen
  \bibfield  {author} {\bibinfo {author} {\bibfnamefont {T.}~\bibnamefont
  {Yanai}}, \bibinfo {author} {\bibfnamefont {G.~I.}\ \bibnamefont {Fann}},
  \bibinfo {author} {\bibfnamefont {Z.}~\bibnamefont {Gan}}, \bibinfo {author}
  {\bibfnamefont {R.~J.}\ \bibnamefont {Harrison}}, \ and\ \bibinfo {author}
  {\bibfnamefont {G.}~\bibnamefont {Beylkin}},\ }\href
  {http://eutils.ncbi.nlm.nih.gov/entrez/eutils/elink.fcgi?dbfrom=pubmed&id=15291596&retmode=ref&cmd=prlinks}
  {\bibfield  {journal} {\bibinfo  {journal} {J. Chem. Phys.}\ }\textbf
  {\bibinfo {volume} {121}},\ \bibinfo {pages} {2866} (\bibinfo {year}
  {2004}{\natexlab{b}})}\BibitemShut {NoStop}%
\bibitem [{\citenamefont {Vosko}\ \emph {et~al.}(1980)\citenamefont {Vosko},
  \citenamefont {Wilk},\ and\ \citenamefont {Nusair}}]{svwn5:1980}%
  \BibitemOpen
  \bibfield  {author} {\bibinfo {author} {\bibfnamefont {S.~H.}\ \bibnamefont
  {Vosko}}, \bibinfo {author} {\bibfnamefont {L.}~\bibnamefont {Wilk}}, \ and\
  \bibinfo {author} {\bibfnamefont {M.}~\bibnamefont {Nusair}},\ }\href@noop {}
  {\bibfield  {journal} {\bibinfo  {journal} {Can. J. Phys.}\ }\textbf
  {\bibinfo {volume} {58}},\ \bibinfo {pages} {1200} (\bibinfo {year}
  {1980})}\BibitemShut {NoStop}%
\bibitem [{\citenamefont {Adamo}\ and\ \citenamefont
  {Barone}(1999)}]{pbe0adam}%
  \BibitemOpen
  \bibfield  {author} {\bibinfo {author} {\bibfnamefont {C.}~\bibnamefont
  {Adamo}}\ and\ \bibinfo {author} {\bibfnamefont {V.}~\bibnamefont {Barone}},\
  }\href@noop {} {\bibfield  {journal} {\bibinfo  {journal} {J. Chem. Phys.}\
  }\textbf {\bibinfo {volume} {110}},\ \bibinfo {pages} {6158} (\bibinfo {year}
  {1999})}\BibitemShut {NoStop}%
\bibitem [{\citenamefont {Paier}\ \emph {et~al.}(2005)\citenamefont {Paier},
  \citenamefont {Hirschl}, \citenamefont {Marsman},\ and\ \citenamefont
  {Kresse}}]{paier2005perdew}%
  \BibitemOpen
  \bibfield  {author} {\bibinfo {author} {\bibfnamefont {J.}~\bibnamefont
  {Paier}}, \bibinfo {author} {\bibfnamefont {R.}~\bibnamefont {Hirschl}},
  \bibinfo {author} {\bibfnamefont {M.}~\bibnamefont {Marsman}}, \ and\
  \bibinfo {author} {\bibfnamefont {G.}~\bibnamefont {Kresse}},\ }\href@noop {}
  {\bibfield  {journal} {\bibinfo  {journal} {J. Chem. Phys.}\ }\textbf
  {\bibinfo {volume} {122}},\ \bibinfo {pages} {234102} (\bibinfo {year}
  {2005})}\BibitemShut {NoStop}%
\bibitem [{\citenamefont {Medvedev}\ \emph {et~al.}(2017)\citenamefont
  {Medvedev}, \citenamefont {Bushmarinov}, \citenamefont {Sun}, \citenamefont
  {Perdew},\ and\ \citenamefont {Lyssenko}}]{Medvedev:2017dj}%
  \BibitemOpen
  \bibfield  {author} {\bibinfo {author} {\bibfnamefont {M.~G.}\ \bibnamefont
  {Medvedev}}, \bibinfo {author} {\bibfnamefont {I.~S.}\ \bibnamefont
  {Bushmarinov}}, \bibinfo {author} {\bibfnamefont {J.}~\bibnamefont {Sun}},
  \bibinfo {author} {\bibfnamefont {J.~P.}\ \bibnamefont {Perdew}}, \ and\
  \bibinfo {author} {\bibfnamefont {K.~A.}\ \bibnamefont {Lyssenko}},\ }\href
  {\doibase 10.1126/science.aah5975} {\bibfield  {journal} {\bibinfo  {journal}
  {Science}\ }\textbf {\bibinfo {volume} {355}},\ \bibinfo {pages} {49}
  (\bibinfo {year} {2017})}\BibitemShut {NoStop}%
\bibitem [{\citenamefont {Pask}\ and\ \citenamefont {Sterne}(2005)}]{pask}%
  \BibitemOpen
  \bibfield  {author} {\bibinfo {author} {\bibfnamefont {J.~E.}\ \bibnamefont
  {Pask}}\ and\ \bibinfo {author} {\bibfnamefont {P.~A.}\ \bibnamefont
  {Sterne}},\ }\href {http://stacks.iop.org/0965-0393/13/i=3/a=R01} {\bibfield
  {journal} {\bibinfo  {journal} {Modelling and Simulation in Materials Science
  and Engineering}\ }\textbf {\bibinfo {volume} {13}},\ \bibinfo {pages} {R71}
  (\bibinfo {year} {2005})}\BibitemShut {NoStop}%
\bibitem [{\citenamefont {Losilla}\ and\ \citenamefont
  {Sundholm}(2012)}]{Losilla:2012bu}%
  \BibitemOpen
  \bibfield  {author} {\bibinfo {author} {\bibfnamefont {S.~A.}\ \bibnamefont
  {Losilla}}\ and\ \bibinfo {author} {\bibfnamefont {D.}~\bibnamefont
  {Sundholm}},\ }\href {\doibase 10.1063/1.4721386} {\bibfield  {journal}
  {\bibinfo  {journal} {J. Chem. Phys.}\ }\textbf {\bibinfo {volume} {136}},\
  \bibinfo {pages} {214104} (\bibinfo {year} {2012})}\BibitemShut {NoStop}%
\bibitem [{\citenamefont {Toivanen}\ \emph {et~al.}(2015)\citenamefont
  {Toivanen}, \citenamefont {Losilla},\ and\ \citenamefont
  {Sundholm}}]{Toivanen:2015hp}%
  \BibitemOpen
  \bibfield  {author} {\bibinfo {author} {\bibfnamefont {E.~A.}\ \bibnamefont
  {Toivanen}}, \bibinfo {author} {\bibfnamefont {S.~A.}\ \bibnamefont
  {Losilla}}, \ and\ \bibinfo {author} {\bibfnamefont {D.}~\bibnamefont
  {Sundholm}},\ }\href {\doibase 10.1039/c5cp01173f} {\bibfield  {journal}
  {\bibinfo  {journal} {Phys. Chem. Chem. Phys.}\ }\textbf {\bibinfo {volume}
  {17}},\ \bibinfo {pages} {31480} (\bibinfo {year} {2015})}\BibitemShut
  {NoStop}%
\bibitem [{\citenamefont {Parkkinen}\ \emph {et~al.}(2017)\citenamefont
  {Parkkinen}, \citenamefont {Losilla}, \citenamefont {Solala}, \citenamefont
  {Toivanen}, \citenamefont {Xu},\ and\ \citenamefont
  {Sundholm}}]{doi:10.1021/acs.jctc.6b01207}%
  \BibitemOpen
  \bibfield  {author} {\bibinfo {author} {\bibfnamefont {P.}~\bibnamefont
  {Parkkinen}}, \bibinfo {author} {\bibfnamefont {S.~A.}\ \bibnamefont
  {Losilla}}, \bibinfo {author} {\bibfnamefont {E.}~\bibnamefont {Solala}},
  \bibinfo {author} {\bibfnamefont {E.~A.}\ \bibnamefont {Toivanen}}, \bibinfo
  {author} {\bibfnamefont {W.-H.}\ \bibnamefont {Xu}}, \ and\ \bibinfo {author}
  {\bibfnamefont {D.}~\bibnamefont {Sundholm}},\ }\href
  {http://dx.doi.org/10.1021/acs.jctc.6b01207} {\bibfield  {journal} {\bibinfo
  {journal} {J. Chem. Theory and Comput.}\ } (\bibinfo {year} {2017})},\
  \bibinfo {note} {dOI:10.1021/acs.jctc.6b01207},\ \Eprint
  {http://arxiv.org/abs/http://dx.doi.org/10.1021/acs.jctc.6b01207}
  {http://dx.doi.org/10.1021/acs.jctc.6b01207} \BibitemShut {NoStop}%
\bibitem [{\citenamefont {Lin}\ \emph {et~al.}(2012)\citenamefont {Lin},
  \citenamefont {Lu}, \citenamefont {Ying},\ and\ \citenamefont
  {E}}]{galerkin}%
  \BibitemOpen
  \bibfield  {author} {\bibinfo {author} {\bibfnamefont {L.}~\bibnamefont
  {Lin}}, \bibinfo {author} {\bibfnamefont {J.}~\bibnamefont {Lu}}, \bibinfo
  {author} {\bibfnamefont {L.}~\bibnamefont {Ying}}, \ and\ \bibinfo {author}
  {\bibfnamefont {W.}~\bibnamefont {E}},\ }\href {\doibase
  http://dx.doi.org/10.1016/j.jcp.2011.11.032} {\bibfield  {journal} {\bibinfo
  {journal} {J. Comput. Phys.}\ }\textbf {\bibinfo {volume} {231}},\ \bibinfo
  {pages} {2140 } (\bibinfo {year} {2012})}\BibitemShut {NoStop}%
\bibitem [{\citenamefont {Frediani}\ and\ \citenamefont
  {Sundholm}(2015)}]{frediani2015real}%
  \BibitemOpen
  \bibfield  {author} {\bibinfo {author} {\bibfnamefont {L.}~\bibnamefont
  {Frediani}}\ and\ \bibinfo {author} {\bibfnamefont {D.}~\bibnamefont
  {Sundholm}},\ }\href@noop {} {\bibfield  {journal} {\bibinfo  {journal}
  {Phys. Chem. Chem. Phys.}\ }\textbf {\bibinfo {volume} {17}},\ \bibinfo
  {pages} {31357} (\bibinfo {year} {2015})}\BibitemShut {NoStop}%
\bibitem [{\citenamefont {Sundholm}\ and\ \citenamefont
  {Olsen}(1994)}]{Sundholm:1994hj}%
  \BibitemOpen
  \bibfield  {author} {\bibinfo {author} {\bibfnamefont {D.}~\bibnamefont
  {Sundholm}}\ and\ \bibinfo {author} {\bibfnamefont {J.}~\bibnamefont
  {Olsen}},\ }\href {\doibase 10.1103/PhysRevA.49.3453} {\bibfield  {journal}
  {\bibinfo  {journal} {Phys. Rev. A}\ }\textbf {\bibinfo {volume} {49}},\
  \bibinfo {pages} {3453} (\bibinfo {year} {1994})}\BibitemShut {NoStop}%
\bibitem [{\citenamefont {Alpert}\ \emph {et~al.}(1993)\citenamefont {Alpert},
  \citenamefont {Beylkin},\ and\ \citenamefont {Coifman}}]{Alpert:1993wd}%
  \BibitemOpen
  \bibfield  {author} {\bibinfo {author} {\bibfnamefont {B.~K.}\ \bibnamefont
  {Alpert}}, \bibinfo {author} {\bibfnamefont {G.}~\bibnamefont {Beylkin}}, \
  and\ \bibinfo {author} {\bibfnamefont {R.}~\bibnamefont {Coifman}},\
  }\href@noop {} {\bibfield  {journal} {\bibinfo  {journal} {SIAM J. Sci. Stat.
  Comput.}\ }\textbf {\bibinfo {volume} {14}},\ \bibinfo {pages} {159}
  (\bibinfo {year} {1993})}\BibitemShut {NoStop}%
\bibitem [{\citenamefont {Alpert}(1999)}]{Alpert:1999hv}%
  \BibitemOpen
  \bibfield  {author} {\bibinfo {author} {\bibfnamefont {B.~K.}\ \bibnamefont
  {Alpert}},\ }\href {\doibase 10.1137/0524016} {\bibfield  {journal} {\bibinfo
   {journal} {SIAM J. Math. Analysis}\ }\textbf {\bibinfo {volume} {24}},\
  \bibinfo {pages} {246} (\bibinfo {year} {1999})}\BibitemShut {NoStop}%
\bibitem [{\citenamefont {Beylkin}\ and\ \citenamefont
  {Mohlenkamp}(2002)}]{Beylkin:2002vla}%
  \BibitemOpen
  \bibfield  {author} {\bibinfo {author} {\bibfnamefont {G.}~\bibnamefont
  {Beylkin}}\ and\ \bibinfo {author} {\bibfnamefont {M.}~\bibnamefont
  {Mohlenkamp}},\ }\href@noop {} {\bibfield  {journal} {\bibinfo  {journal}
  {Proc. Natl. Acad. Sci. USA}\ }\textbf {\bibinfo {volume} {99}},\ \bibinfo
  {pages} {10246} (\bibinfo {year} {2002})}\BibitemShut {NoStop}%
\bibitem [{\citenamefont {Beylkin}\ \emph {et~al.}(2007)\citenamefont
  {Beylkin}, \citenamefont {Cramer}, \citenamefont {Fann},\ and\ \citenamefont
  {Harrison}}]{Beylkin:2007gv}%
  \BibitemOpen
  \bibfield  {author} {\bibinfo {author} {\bibfnamefont {G.}~\bibnamefont
  {Beylkin}}, \bibinfo {author} {\bibfnamefont {R.}~\bibnamefont {Cramer}},
  \bibinfo {author} {\bibfnamefont {G.}~\bibnamefont {Fann}}, \ and\ \bibinfo
  {author} {\bibfnamefont {R.~J.}\ \bibnamefont {Harrison}},\ }\href {\doibase
  10.1016/j.acha.2007.01.001} {\bibfield  {journal} {\bibinfo  {journal} {Appl.
  Comput. Harmon. A.}\ }\textbf {\bibinfo {volume} {23}},\ \bibinfo {pages}
  {235} (\bibinfo {year} {2007})}\BibitemShut {NoStop}%
\bibitem [{\citenamefont {Fann}\ \emph {et~al.}(2004)\citenamefont {Fann},
  \citenamefont {Beylkin}, \citenamefont {Harrison},\ and\ \citenamefont
  {Jordan}}]{Fann:2004ub}%
  \BibitemOpen
  \bibfield  {author} {\bibinfo {author} {\bibfnamefont {G.}~\bibnamefont
  {Fann}}, \bibinfo {author} {\bibfnamefont {G.}~\bibnamefont {Beylkin}},
  \bibinfo {author} {\bibfnamefont {R.~J.}\ \bibnamefont {Harrison}}, \ and\
  \bibinfo {author} {\bibfnamefont {K.~E.}\ \bibnamefont {Jordan}},\ }\href
  {http://ieeexplore.ieee.org/xpls/abs_all.jsp?arnumber=5388865} {\bibfield
  {journal} {\bibinfo  {journal} {IBM J. Research and Development}\ }\textbf
  {\bibinfo {volume} {48}},\ \bibinfo {pages} {161} (\bibinfo {year}
  {2004})}\BibitemShut {NoStop}%
\bibitem [{\citenamefont {Beylkin}(2005)}]{Beylkin:2005wz}%
  \BibitemOpen
  \bibfield  {author} {\bibinfo {author} {\bibfnamefont {G.}~\bibnamefont
  {Beylkin}},\ }\href@noop {} {\bibfield  {journal} {\bibinfo  {journal} {Appl.
  Comput. Harmon. A.}\ }\textbf {\bibinfo {volume} {19}},\ \bibinfo {pages}
  {17} (\bibinfo {year} {2005})}\BibitemShut {NoStop}%
\bibitem [{\citenamefont {Gines}\ \emph {et~al.}(1998)\citenamefont {Gines},
  \citenamefont {Beylkin},\ and\ \citenamefont {Dunn}}]{Gines:1998uw}%
  \BibitemOpen
  \bibfield  {author} {\bibinfo {author} {\bibfnamefont {D.}~\bibnamefont
  {Gines}}, \bibinfo {author} {\bibfnamefont {G.}~\bibnamefont {Beylkin}}, \
  and\ \bibinfo {author} {\bibfnamefont {J.}~\bibnamefont {Dunn}},\ }\href@noop
  {} {\bibfield  {journal} {\bibinfo  {journal} {Appl. Comput. Harmon. A.}\
  }\textbf {\bibinfo {volume} {5}},\ \bibinfo {pages} {156} (\bibinfo {year}
  {1998})}\BibitemShut {NoStop}%
\bibitem [{\citenamefont {Beylkin}\ \emph {et~al.}(2008)\citenamefont
  {Beylkin}, \citenamefont {Cheruvu},\ and\ \citenamefont
  {Perez}}]{Beylkin:2008im}%
  \BibitemOpen
  \bibfield  {author} {\bibinfo {author} {\bibfnamefont {G.}~\bibnamefont
  {Beylkin}}, \bibinfo {author} {\bibfnamefont {V.}~\bibnamefont {Cheruvu}}, \
  and\ \bibinfo {author} {\bibfnamefont {F.}~\bibnamefont {Perez}},\ }\href
  {\doibase 10.1016/j.acha.2007.08.001} {\bibfield  {journal} {\bibinfo
  {journal} {Appl. Comput. Harmon. A.}\ }\textbf {\bibinfo {volume} {24}},\
  \bibinfo {pages} {354} (\bibinfo {year} {2008})}\BibitemShut {NoStop}%
\bibitem [{\citenamefont {Kalos}(1962)}]{Kalos:1962uj}%
  \BibitemOpen
  \bibfield  {author} {\bibinfo {author} {\bibfnamefont {M.~H.}\ \bibnamefont
  {Kalos}},\ }\href {http://prola.aps.org/abstract/PR/v128/i4/p1791_1}
  {\bibfield  {journal} {\bibinfo  {journal} {Phys. Rev.}\ }\textbf {\bibinfo
  {volume} {128}},\ \bibinfo {pages} {1791} (\bibinfo {year}
  {1962})}\BibitemShut {NoStop}%
\bibitem [{\citenamefont {Harrison}(2004)}]{Harrison:2004gd}%
  \BibitemOpen
  \bibfield  {author} {\bibinfo {author} {\bibfnamefont {R.~J.}\ \bibnamefont
  {Harrison}},\ }\href {\doibase 10.1002/jcc.10108} {\bibfield  {journal}
  {\bibinfo  {journal} {J. Comput. Chem.}\ }\textbf {\bibinfo {volume} {25}},\
  \bibinfo {pages} {328} (\bibinfo {year} {2004})}\BibitemShut {NoStop}%
\bibitem [{\citenamefont {Yanai}\ \emph {et~al.}(2005)\citenamefont {Yanai},
  \citenamefont {Harrison},\ and\ \citenamefont {Handy}}]{Yanai:2005en}%
  \BibitemOpen
  \bibfield  {author} {\bibinfo {author} {\bibfnamefont {T.}~\bibnamefont
  {Yanai}}, \bibinfo {author} {\bibfnamefont {R.~J.}\ \bibnamefont {Harrison}},
  \ and\ \bibinfo {author} {\bibfnamefont {N.~C.}\ \bibnamefont {Handy}},\
  }\href {\doibase 10.1080/00268970412331319236} {\bibfield  {journal}
  {\bibinfo  {journal} {Mol. Phys.}\ }\textbf {\bibinfo {volume} {103}},\
  \bibinfo {pages} {413} (\bibinfo {year} {2005})}\BibitemShut {NoStop}%
\bibitem [{\citenamefont {Yanai}\ \emph {et~al.}(2015)\citenamefont {Yanai},
  \citenamefont {Fann}, \citenamefont {Beylkin},\ and\ \citenamefont
  {Harrison}}]{Yanai:2015gb}%
  \BibitemOpen
  \bibfield  {author} {\bibinfo {author} {\bibfnamefont {T.}~\bibnamefont
  {Yanai}}, \bibinfo {author} {\bibfnamefont {G.~I.}\ \bibnamefont {Fann}},
  \bibinfo {author} {\bibfnamefont {G.}~\bibnamefont {Beylkin}}, \ and\
  \bibinfo {author} {\bibfnamefont {R.~J.}\ \bibnamefont {Harrison}},\ }\href
  {\doibase 10.1039/c4cp05821f} {\bibfield  {journal} {\bibinfo  {journal}
  {Phys. Chem. Chem. Phys.}\ }\textbf {\bibinfo {volume} {17}},\ \bibinfo
  {pages} {31405} (\bibinfo {year} {2015})}\BibitemShut {NoStop}%
\bibitem [{\citenamefont {Kottmann}\ \emph {et~al.}(2015)\citenamefont
  {Kottmann}, \citenamefont {H{\"o}fener},\ and\ \citenamefont
  {Bischoff}}]{Kottmann:2015gc}%
  \BibitemOpen
  \bibfield  {author} {\bibinfo {author} {\bibfnamefont {J.~S.}\ \bibnamefont
  {Kottmann}}, \bibinfo {author} {\bibfnamefont {S.}~\bibnamefont
  {H{\"o}fener}}, \ and\ \bibinfo {author} {\bibfnamefont {F.~A.}\ \bibnamefont
  {Bischoff}},\ }\href {\doibase 10.1039/c5cp00345h} {\bibfield  {journal}
  {\bibinfo  {journal} {Phys. Chem. Chem. Phys.}\ }\textbf {\bibinfo {volume}
  {17}},\ \bibinfo {pages} {31453} (\bibinfo {year} {2015})}\BibitemShut
  {NoStop}%
\bibitem [{\citenamefont {Sekino}\ \emph {et~al.}(2008)\citenamefont {Sekino},
  \citenamefont {Maeda}, \citenamefont {Yanai},\ and\ \citenamefont
  {Harrison}}]{Sekino:2008ef}%
  \BibitemOpen
  \bibfield  {author} {\bibinfo {author} {\bibfnamefont {H.}~\bibnamefont
  {Sekino}}, \bibinfo {author} {\bibfnamefont {Y.}~\bibnamefont {Maeda}},
  \bibinfo {author} {\bibfnamefont {T.}~\bibnamefont {Yanai}}, \ and\ \bibinfo
  {author} {\bibfnamefont {R.~J.}\ \bibnamefont {Harrison}},\ }\href {\doibase
  10.1063/1.2955730} {\bibfield  {journal} {\bibinfo  {journal} {J. Chem.
  Phys.}\ }\textbf {\bibinfo {volume} {129}},\ \bibinfo {pages} {034111}
  (\bibinfo {year} {2008})}\BibitemShut {NoStop}%
\bibitem [{\citenamefont {Sekino}\ \emph {et~al.}(2012)\citenamefont {Sekino},
  \citenamefont {Yokoi},\ and\ \citenamefont {Harrison}}]{Sekino:2012ge}%
  \BibitemOpen
  \bibfield  {author} {\bibinfo {author} {\bibfnamefont {H.}~\bibnamefont
  {Sekino}}, \bibinfo {author} {\bibfnamefont {Y.}~\bibnamefont {Yokoi}}, \
  and\ \bibinfo {author} {\bibfnamefont {R.~J.}\ \bibnamefont {Harrison}},\
  }\href {\doibase 10.1088/1742-6596/352/1/012014} {\bibfield  {journal}
  {\bibinfo  {journal} {J. Phys.: Conf. Series}\ }\textbf {\bibinfo {volume}
  {352}},\ \bibinfo {pages} {012014} (\bibinfo {year} {2012})}\BibitemShut
  {NoStop}%
\bibitem [{\citenamefont {Jensen}\ \emph {et~al.}(2016)\citenamefont {Jensen},
  \citenamefont {Fl{\aa}}, \citenamefont {Jonsson}, \citenamefont {Monstad},
  \citenamefont {Ruud},\ and\ \citenamefont {Frediani}}]{Jensen:2016jy}%
  \BibitemOpen
  \bibfield  {author} {\bibinfo {author} {\bibfnamefont {S.~R.}\ \bibnamefont
  {Jensen}}, \bibinfo {author} {\bibfnamefont {T.}~\bibnamefont {Fl{\aa}}},
  \bibinfo {author} {\bibfnamefont {D.}~\bibnamefont {Jonsson}}, \bibinfo
  {author} {\bibfnamefont {R.~S.}\ \bibnamefont {Monstad}}, \bibinfo {author}
  {\bibfnamefont {K.}~\bibnamefont {Ruud}}, \ and\ \bibinfo {author}
  {\bibfnamefont {L.}~\bibnamefont {Frediani}},\ }\href {\doibase
  10.1039/C6CP01294A} {\bibfield  {journal} {\bibinfo  {journal} {Phys. Chem.
  Chem. Phys.}\ }\textbf {\bibinfo {volume} {18}},\ \bibinfo {pages} {21145}
  (\bibinfo {year} {2016})}\BibitemShut {NoStop}%
\bibitem [{Note1()}]{Note1}%
  \BibitemOpen
  \bibinfo {note} {For the singularity of the nuclear potential, a simple
  smoothing is employed, but its effect on the accuracy is easily controlled by
  a single parameter~\cite {Harrison:2004vua}.}\BibitemShut {Stop}%
\bibitem [{Note2()}]{Note2}%
  \BibitemOpen
  \bibinfo {note} {Displayed as spin-unpolarized for clarity. Extension to
  spin-\ac {DFT} is fairly straightforward.}\BibitemShut {Stop}%
\bibitem [{\citenamefont {Marques}\ \emph {et~al.}(2012)\citenamefont
  {Marques}, \citenamefont {Oliveira},\ and\ \citenamefont
  {Burnus}}]{marques2012libxc}%
  \BibitemOpen
  \bibfield  {author} {\bibinfo {author} {\bibfnamefont {M.~A.}\ \bibnamefont
  {Marques}}, \bibinfo {author} {\bibfnamefont {M.~J.}\ \bibnamefont
  {Oliveira}}, \ and\ \bibinfo {author} {\bibfnamefont {T.}~\bibnamefont
  {Burnus}},\ }\href@noop {} {\bibfield  {journal} {\bibinfo  {journal}
  {Comput. Phys. Comm.}\ }\textbf {\bibinfo {volume} {183}},\ \bibinfo {pages}
  {2272} (\bibinfo {year} {2012})}\BibitemShut {NoStop}%
\bibitem [{\citenamefont {Ekstr{\"o}m}\ \emph {et~al.}(2010)\citenamefont
  {Ekstr{\"o}m}, \citenamefont {Visscher}, \citenamefont {Bast}, \citenamefont
  {Thorvaldsen},\ and\ \citenamefont {Ruud}}]{Ekstrom:2010bz}%
  \BibitemOpen
  \bibfield  {author} {\bibinfo {author} {\bibfnamefont {U.}~\bibnamefont
  {Ekstr{\"o}m}}, \bibinfo {author} {\bibfnamefont {L.}~\bibnamefont
  {Visscher}}, \bibinfo {author} {\bibfnamefont {R.}~\bibnamefont {Bast}},
  \bibinfo {author} {\bibfnamefont {A.~J.}\ \bibnamefont {Thorvaldsen}}, \ and\
  \bibinfo {author} {\bibfnamefont {K.}~\bibnamefont {Ruud}},\ }\href {\doibase
  10.1021/ct100117s} {\bibfield  {journal} {\bibinfo  {journal} {J. Chem.
  Theory Comput.}\ }\textbf {\bibinfo {volume} {6}},\ \bibinfo {pages} {1971}
  (\bibinfo {year} {2010})}\BibitemShut {NoStop}%
\bibitem [{\citenamefont {Alpert}\ \emph {et~al.}(2002)\citenamefont {Alpert},
  \citenamefont {Beylkin}, \citenamefont {Gines},\ and\ \citenamefont
  {Vozovoi}}]{Alpert:1999tk}%
  \BibitemOpen
  \bibfield  {author} {\bibinfo {author} {\bibfnamefont {B.~K.}\ \bibnamefont
  {Alpert}}, \bibinfo {author} {\bibfnamefont {G.}~\bibnamefont {Beylkin}},
  \bibinfo {author} {\bibfnamefont {D.}~\bibnamefont {Gines}}, \ and\ \bibinfo
  {author} {\bibfnamefont {L.}~\bibnamefont {Vozovoi}},\ }\href
  {http://www.sciencedirect.com/science/article/pii/S0021999102971603}
  {\bibfield  {journal} {\bibinfo  {journal} {J. Comput. Phys.}\ }\textbf
  {\bibinfo {volume} {182}},\ \bibinfo {pages} {149} (\bibinfo {year}
  {2002})}\BibitemShut {NoStop}%
\bibitem [{Note3()}]{Note3}%
  \BibitemOpen
  \bibinfo {note} {This update is exact, provided that the orbital update comes
  directly from the application of the \ac {BSH} operator defined by the
  previous (not necessarily exact) eigenvalue: $\Delta \protect \ensuremath
  {\varphi }^n = -2\protect \mathaccentV {hat}05E{G}^n_\mu \left [\protect
  \mathaccentV {hat}05E{V}^n\protect \ensuremath {\varphi }^n\right ] -
  \protect \ensuremath {\varphi }^n$. Generalizations can be made for multiple
  orbitals.}\BibitemShut {Stop}%
\bibitem [{Note4()}]{Note4}%
  \BibitemOpen
  \bibinfo {note} {The grid is constructed such that it holds both the density
  and its gradient within the requested accuracy.}\BibitemShut {Stop}%
\bibitem [{mrc(2016)}]{mrchem}%
  \BibitemOpen
  \href@noop {} {\enquote {\bibinfo {title} {Multiresolution chemistry (mrchem)
  program package.}}\ }\bibinfo {howpublished} {See
  http://mrchemdoc.readthedocs.org/en/latest/} (\bibinfo {year}
  {2016})\BibitemShut {NoStop}%
\bibitem [{\citenamefont {Feller}(1996)}]{feller1996role}%
  \BibitemOpen
  \bibfield  {author} {\bibinfo {author} {\bibfnamefont {D.}~\bibnamefont
  {Feller}},\ }\href@noop {} {\bibfield  {journal} {\bibinfo  {journal} {J.
  Comput. Chem.}\ }\textbf {\bibinfo {volume} {17}},\ \bibinfo {pages} {1571}
  (\bibinfo {year} {1996})}\BibitemShut {NoStop}%
\bibitem [{\citenamefont {Schuchardt}\ \emph {et~al.}(2007)\citenamefont
  {Schuchardt}, \citenamefont {Didier}, \citenamefont {Elsethagen},
  \citenamefont {Sun}, \citenamefont {Gurumoorthi}, \citenamefont {Chase},
  \citenamefont {Li},\ and\ \citenamefont {Windus}}]{schuchardt2007basis}%
  \BibitemOpen
  \bibfield  {author} {\bibinfo {author} {\bibfnamefont {K.~L.}\ \bibnamefont
  {Schuchardt}}, \bibinfo {author} {\bibfnamefont {B.~T.}\ \bibnamefont
  {Didier}}, \bibinfo {author} {\bibfnamefont {T.}~\bibnamefont {Elsethagen}},
  \bibinfo {author} {\bibfnamefont {L.}~\bibnamefont {Sun}}, \bibinfo {author}
  {\bibfnamefont {V.}~\bibnamefont {Gurumoorthi}}, \bibinfo {author}
  {\bibfnamefont {J.}~\bibnamefont {Chase}}, \bibinfo {author} {\bibfnamefont
  {J.}~\bibnamefont {Li}}, \ and\ \bibinfo {author} {\bibfnamefont {T.~L.}\
  \bibnamefont {Windus}},\ }\href@noop {} {\bibfield  {journal} {\bibinfo
  {journal} {J. Chem. Information and Modeling}\ }\textbf {\bibinfo {volume}
  {47}},\ \bibinfo {pages} {1045} (\bibinfo {year} {2007})}\BibitemShut
  {NoStop}%
\bibitem [{\citenamefont {Valiev}\ \emph {et~al.}(2010)\citenamefont {Valiev},
  \citenamefont {Bylaska}, \citenamefont {Govind}, \citenamefont {Kowalski},
  \citenamefont {Straatsma}, \citenamefont {van Dam}, \citenamefont {Wang},
  \citenamefont {Nieplocha}, \citenamefont {Apra}, \citenamefont {Windus},\
  and\ \citenamefont {de~Jong}}]{nwchem}%
  \BibitemOpen
  \bibfield  {author} {\bibinfo {author} {\bibfnamefont {M.}~\bibnamefont
  {Valiev}}, \bibinfo {author} {\bibfnamefont {E.}~\bibnamefont {Bylaska}},
  \bibinfo {author} {\bibfnamefont {N.}~\bibnamefont {Govind}}, \bibinfo
  {author} {\bibfnamefont {K.}~\bibnamefont {Kowalski}}, \bibinfo {author}
  {\bibfnamefont {T.}~\bibnamefont {Straatsma}}, \bibinfo {author}
  {\bibfnamefont {H.}~\bibnamefont {van Dam}}, \bibinfo {author} {\bibfnamefont
  {D.}~\bibnamefont {Wang}}, \bibinfo {author} {\bibfnamefont {J.}~\bibnamefont
  {Nieplocha}}, \bibinfo {author} {\bibfnamefont {E.}~\bibnamefont {Apra}},
  \bibinfo {author} {\bibfnamefont {T.}~\bibnamefont {Windus}}, \ and\ \bibinfo
  {author} {\bibfnamefont {W.}~\bibnamefont {de~Jong}},\ }\href@noop {}
  {\bibfield  {journal} {\bibinfo  {journal} {Comput. Phys. Comm.}\ }\textbf
  {\bibinfo {volume} {181}},\ \bibinfo {pages} {1477} (\bibinfo {year}
  {2010})}\BibitemShut {NoStop}%
\bibitem [{\citenamefont {Delley}(1990)}]{delley1990all}%
  \BibitemOpen
  \bibfield  {author} {\bibinfo {author} {\bibfnamefont {B.}~\bibnamefont
  {Delley}},\ }\href@noop {} {\bibfield  {journal} {\bibinfo  {journal} {J.
  Chem. Phys.}\ }\textbf {\bibinfo {volume} {92}},\ \bibinfo {pages} {508}
  (\bibinfo {year} {1990})}\BibitemShut {NoStop}%
\bibitem [{\citenamefont {Blum}\ \emph {et~al.}(2009)\citenamefont {Blum},
  \citenamefont {Gehrke}, \citenamefont {Hanke}, \citenamefont {Havu},
  \citenamefont {Havu}, \citenamefont {Ren}, \citenamefont {Reuter},\ and\
  \citenamefont {Scheffler}}]{fhiaims}%
  \BibitemOpen
  \bibfield  {author} {\bibinfo {author} {\bibfnamefont {V.}~\bibnamefont
  {Blum}}, \bibinfo {author} {\bibfnamefont {R.}~\bibnamefont {Gehrke}},
  \bibinfo {author} {\bibfnamefont {F.}~\bibnamefont {Hanke}}, \bibinfo
  {author} {\bibfnamefont {P.}~\bibnamefont {Havu}}, \bibinfo {author}
  {\bibfnamefont {V.}~\bibnamefont {Havu}}, \bibinfo {author} {\bibfnamefont
  {X.}~\bibnamefont {Ren}}, \bibinfo {author} {\bibfnamefont {K.}~\bibnamefont
  {Reuter}}, \ and\ \bibinfo {author} {\bibfnamefont {M.}~\bibnamefont
  {Scheffler}},\ }\href@noop {} {\bibfield  {journal} {\bibinfo  {journal}
  {Comput. Phys. Comm.}\ }\textbf {\bibinfo {volume} {180}},\ \bibinfo {pages}
  {2175} (\bibinfo {year} {2009})}\BibitemShut {NoStop}%
\bibitem [{\citenamefont {Ren}\ \emph {et~al.}(2012)\citenamefont {Ren},
  \citenamefont {Rinke}, \citenamefont {Blum}, \citenamefont {Wieferink},
  \citenamefont {Tkatchenko}, \citenamefont {Sanfilippo}, \citenamefont
  {Reuter},\ and\ \citenamefont {Scheffler}}]{1367-2630-14-5-053020}%
  \BibitemOpen
  \bibfield  {author} {\bibinfo {author} {\bibfnamefont {X.}~\bibnamefont
  {Ren}}, \bibinfo {author} {\bibfnamefont {P.}~\bibnamefont {Rinke}}, \bibinfo
  {author} {\bibfnamefont {V.}~\bibnamefont {Blum}}, \bibinfo {author}
  {\bibfnamefont {J.}~\bibnamefont {Wieferink}}, \bibinfo {author}
  {\bibfnamefont {A.}~\bibnamefont {Tkatchenko}}, \bibinfo {author}
  {\bibfnamefont {A.}~\bibnamefont {Sanfilippo}}, \bibinfo {author}
  {\bibfnamefont {K.}~\bibnamefont {Reuter}}, \ and\ \bibinfo {author}
  {\bibfnamefont {M.}~\bibnamefont {Scheffler}},\ }\href
  {http://stacks.iop.org/1367-2630/14/i=5/a=053020} {\bibfield  {journal}
  {\bibinfo  {journal} {New Journal of Physics}\ }\textbf {\bibinfo {volume}
  {14}},\ \bibinfo {pages} {053020} (\bibinfo {year} {2012})}\BibitemShut
  {NoStop}%
\bibitem [{elk(2015)}]{elk}%
  \BibitemOpen
  \href@noop {} {\enquote {\bibinfo {title} {An all-electron full-potential
  linearised augmented-plane wave (fp-lapw) code.}}\ }\bibinfo {howpublished}
  {See http://http://elk.sourceforge.net/} (\bibinfo {year} {2015})\BibitemShut
  {NoStop}%
\bibitem [{\citenamefont {Jensen}\ \emph {et~al.}(2017)\citenamefont {Jensen},
  \citenamefont {Saha}, \citenamefont {Flores-Livas}, \citenamefont {Huhn},
  \citenamefont {Blum}, \citenamefont {Goedecker},\ and\ \citenamefont
  {Frediani}}]{0EM0EL_2017}%
  \BibitemOpen
  \bibfield  {author} {\bibinfo {author} {\bibfnamefont {S.~R.}\ \bibnamefont
  {Jensen}}, \bibinfo {author} {\bibfnamefont {S.}~\bibnamefont {Saha}},
  \bibinfo {author} {\bibfnamefont {J.~A.}\ \bibnamefont {Flores-Livas}},
  \bibinfo {author} {\bibfnamefont {W.}~\bibnamefont {Huhn}}, \bibinfo {author}
  {\bibfnamefont {V.}~\bibnamefont {Blum}}, \bibinfo {author} {\bibfnamefont
  {S.}~\bibnamefont {Goedecker}}, \ and\ \bibinfo {author} {\bibfnamefont
  {L.}~\bibnamefont {Frediani}},\ }\href {\doibase 10.18710/0EM0EL} {\enquote
  {\bibinfo {title} {Gga-pbe and hybrid-pbe0 energies and dipole moments with
  mrchem, fhi-aims, nwchem and elk},}\ } (\bibinfo {year} {2017})\BibitemShut
  {NoStop}%
\bibitem [{\citenamefont {Schmider}\ and\ \citenamefont
  {Becke}(1998)}]{schmider1998optimized}%
  \BibitemOpen
  \bibfield  {author} {\bibinfo {author} {\bibfnamefont {H.~L.}\ \bibnamefont
  {Schmider}}\ and\ \bibinfo {author} {\bibfnamefont {A.~D.}\ \bibnamefont
  {Becke}},\ }\href@noop {} {\bibfield  {journal} {\bibinfo  {journal} {J.
  Chem. Phys.}\ }\textbf {\bibinfo {volume} {108}},\ \bibinfo {pages} {9624}
  (\bibinfo {year} {1998})}\BibitemShut {NoStop}%
\bibitem [{Note5()}]{Note5}%
  \BibitemOpen
  \bibinfo {note} {For Li, Be, Na and Mg the largest basis set is aug-cc-pVQZ
  and has therefore been employed. For the other elements, the corresponding 6Z
  basis is also available, but attempts to employ such a basis led to
  overcompleteness problems, often resulting in energies higher than the 5Z
  results.}\BibitemShut {Stop}%
\bibitem [{Note6()}]{Note6}%
  \BibitemOpen
  \bibinfo {note} {Due to technical reasons in convergence, CH$_3$CH$_2$O was
  excluded from the PBE0 results, while CCH was excluded in both PBE and
  PBE0.}\BibitemShut {Stop}%
\bibitem [{\citenamefont {Bak}\ \emph {et~al.}(2000)\citenamefont {Bak},
  \citenamefont {Gauss}, \citenamefont {Helgaker}, \citenamefont
  {J{\o}rgensen},\ and\ \citenamefont {Olsen}}]{bak2000accuracy}%
  \BibitemOpen
  \bibfield  {author} {\bibinfo {author} {\bibfnamefont {K.~L.}\ \bibnamefont
  {Bak}}, \bibinfo {author} {\bibfnamefont {J.}~\bibnamefont {Gauss}}, \bibinfo
  {author} {\bibfnamefont {T.}~\bibnamefont {Helgaker}}, \bibinfo {author}
  {\bibfnamefont {P.}~\bibnamefont {J{\o}rgensen}}, \ and\ \bibinfo {author}
  {\bibfnamefont {J.}~\bibnamefont {Olsen}},\ }\href@noop {} {\bibfield
  {journal} {\bibinfo  {journal} {Chem. Phys. Lett.}\ }\textbf {\bibinfo
  {volume} {319}},\ \bibinfo {pages} {563} (\bibinfo {year}
  {2000})}\BibitemShut {NoStop}%
\end{thebibliography}%
